\documentclass[11pt]{article}
\usepackage{fullpage}
\usepackage{cite}
\usepackage{graphicx}
\usepackage{amsmath}
\usepackage{amssymb}
\title{An introduction to webs}
\date{}

\begin{document}
\bibliographystyle{utphys}
\newcommand{\msbar}{\ensuremath{\overline{\text{MS}}}}
\newcommand{\DIS}{\ensuremath{\text{DIS}}}
\newcommand{\abar}{\ensuremath{\bar{\alpha}_S}}
\newcommand{\bb}{\ensuremath{\bar{\beta}_0}}
\newcommand{\rc}{\ensuremath{r_{\text{cut}}}}
\newcommand{\Nd}{\ensuremath{N_{\text{d.o.f.}}}}
\newcommand \slsh [1] {\not\!{#1}}
\setlength{\parindent}{0pt}

\titlepage

\vspace*{0.5cm}

\begin{center}
{\Large \bf An Introduction to Webs}

\vspace*{1cm}
\textsc{C.D. White$^{a}$\footnote{Christopher.White@glasgow.ac.uk} } \\

\vspace*{0.5cm} $^a$ School of Physics and Astronomy, Scottish Universities
Physics \\Alliance, University of Glasgow, Glasgow G12 8QQ, Scotland, UK\\

\end{center}

\vspace*{0.5cm}

\begin{abstract}
Webs are sets of Feynman diagrams that contribute to the exponents of
scattering amplitudes, in the kinematic limit in which emitted
radiation is soft. As such, they have a number of phenomenological and
formal applications, and offer tantalising glimpses into the all-order
structure of perturbative quantum field theory. This article is based
on a series of lectures given to graduate students, and aims to
provide a pedagogical introduction to webs. Topics covered include
exponentiation in (non-)abelian gauge theories, the web mixing matrix
formalism for non-abelian gauge theories, and recent progress on the
calculation of web diagrams. Problems are included throughout the
text, to aid understanding.
\end{abstract}

\vspace*{0.5cm}

\section{Introduction}
\label{sec:intro}
One of the purposes of quantum field theory is the calculation of scattering
amplitudes, which are related to the probability for a given particle collision
to occur. Typically, amplitudes are calculated in perturbation theory, and it 
is well known that such calculations are beset by infinities. Some of these
infinities (ultraviolet divergences) can be absorbed into redefinitions of the
Lagrangian parameters defining the theory, a procedure known as 
{\it renormalisation}. In these lectures, we will be concerned with infrared 
(IR) divergences, which correspond to long distance physics.\\

\begin{figure}
\begin{center}
\scalebox{1.0}{\includegraphics{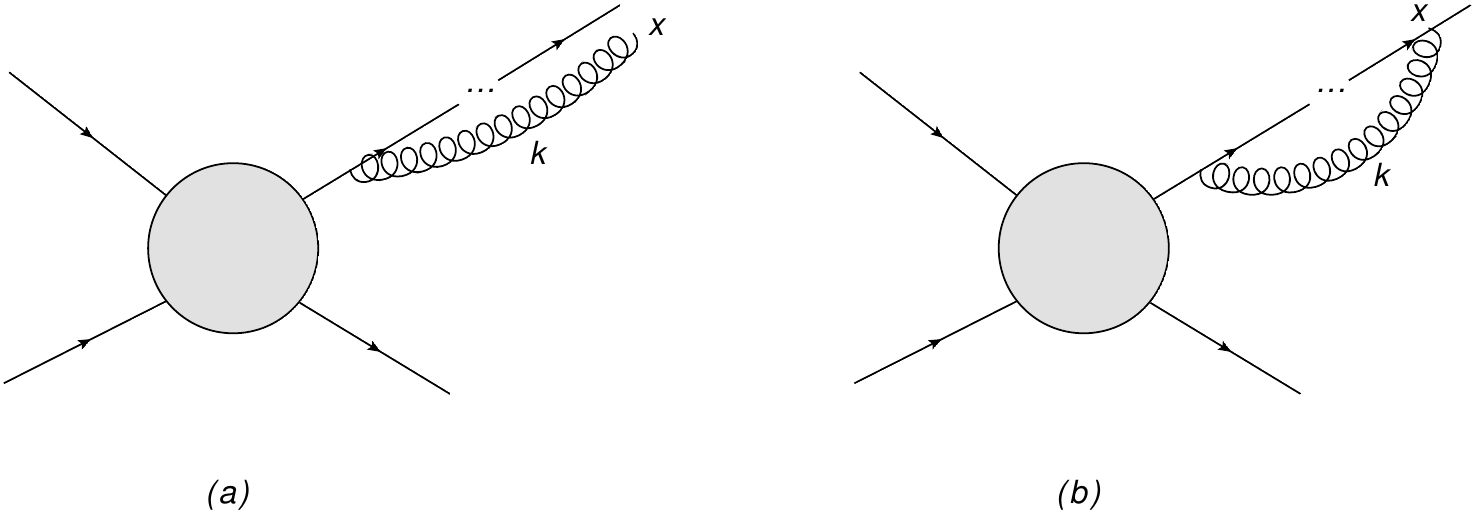}}
\caption{Emission of a gluon of momentum $k$ from an outgoing particle in a
scattering amplitude: (a) real emission; (b) virtual emission.}
\label{fig:longdistance}
\end{center}
\end{figure}

That long distance physics is associated with divergences in perturbation 
theory can be inferred by examining figure~\ref{fig:longdistance}. 
This depicts a 2$\rightarrow$2 scattering process, in which one of the outgoing
particles emits a gluon. This emission may be {\it real}, in which case the 
gluon itself ends up as a final state particle. Alternatively it may be 
{\it virtual}, in which case it joins up again with the outgoing particle. 
These two possibilities are shown in figure~\ref{fig:longdistance}(a) and (b)
respectively. The Feynman rules for emission of a gluon tell us to integrate
over all possible positions for the end of the gluon. That is, there is an 
integral
\begin{displaymath}
\int d^dx\ldots
\end{displaymath}
in $d$ space-time dimensions. Because one integrates out to infinite distance,
there will always be some value of $d$ above which this integral diverges, and 
in fact this is found to be $d=4$. Hence, singularities associated with long
distance physics are a crucial ingredient in practical applications of field
theory. \\

Note that a gluon that extends out to infinite distance has an infinite Compton
wavelength. Thus, by the uncertainty principle, such a gluon has a vanishing
momentum or energy $k^\mu\rightarrow 0$. This is the origin of the phrase
``infrared divergences'' for long distance singularities - if we were 
calculating Feynman diagrams such as those in figure~\ref{fig:longdistance}
in momentum space, the singularities would show up after integrating over the 
emitted gluon momentum $k^\mu$, as contributions from the region where this 
tends to zero in all components. Such gluons are referred to as {\it soft} in 
the literature, as opposed to {\it hard} particles which have nonzero 
momenta. \\

From figure~\ref{fig:longdistance}, we also see that if a gluon has an infinite
Compton wavelength, it is difficult to distinguish between the case of real or
virtual emission, as a virtual gluon will only rejoin an outgoing particle at
infinity. Thus, both real and virtual diagrams will contain infrared 
singularities (after integrating over the gluon position in the real case), and
we expect that the real and virtual contributions for a given soft gluon will
be closely related to each other. This is indeed the case, and in fact
IR divergences cancel upon combining real and virtual graphs. Furthermore, this
is guaranteed, for suitably defined quantities,
to all orders in perturbation theory~\cite{Bloch:1937pw}. However, large
residual contributions to observable quantities remain, which typically take
the form of terms such as
\begin{equation}
\alpha_S^n\frac{\log^mE}{E}
\label{logs}
\end{equation}
involving logarithms of the total energy $E$ carried by soft
gluons. Such terms occur at all orders ${\cal O}(\alpha_S^n)$, and are
highly divergent as $E\rightarrow 0$ such that perturbation theory
breaks down in this regime.  Close to this limit, the logarithmic
terms are not formally singular, but can be numerically large. Then
perturbation theory is no longer sufficiently convergent for a
truncation at e.g. next-to-leading order to be sensible. One must then
sum up the terms of eq.~(\ref{logs}) to all orders in the coupling
constant, a procedure known as {\it resummation}. This is by now a
highly developed field - see
e.g.~\cite{Sterman:1986aj,Catani:1989ne,Contopanagos:1997nh,Korchemsky:1993uz,Korchemsky:1993xv,Forte:2002ni,Becher:2006nr,Schwartz:2007ib,Bauer:2008dt,Chiu:2009mg,Laenen:2008gt}
for several different approaches, and
refs.~\cite{Becher:2014oda,Luisoni:2015xha} for recent review
articles. Implicit in the above discussion is that resummation is
intimately linked with the structure of infrared singularities to all
orders in perturbation theory. \\

Infrared singularities are also of interest in themselves. In
particular, there are a number of recent conjectures regarding IR
divergences in a variety of theories. One example is the so-called
{\it dipole formula} in QCD~\cite{Gardi:2009qi,Gardi:2009zv,
  Becher:2009cu,Becher:2009qa,Becher:2009kw}, an all-order ansatz for
the IR singularities of fixed-angle scattering amplitudes with
massless particles.  There is also interesting recent work on
singularity structures in supersymmetric gauge theories, and their
relationship to similar amplitudes in
(super-)gravity~\cite{Bern:2010yg,Bern:2010ue,Bern:2002kj,
  BoucherVeronneau:2011qv,Naculich:2011my,Naculich:2011pd}. The IR
singularity structure of quantum gravity itself has been studied
in~\cite{Weinberg:1965nx}, and more recently
in~\cite{Naculich:2011ry,White:2011yy,Akhoury:2011kq,Beneke:2012xa}. The
relation between QCD and gravity in the soft limit has been explored
in~\cite{Oxburgh:2012zr}, and used to provide evidence for the {\it
  double copy} conjecture relating both
theories~\cite{Bern:2010yg,Bern:2010ue}.\\

As a result of these investigations, much is already known about the
structure of infrared divergences. Crucially, it is known that
scattering amplitudes {\it factorise} into the following schematic
form~\cite{Mueller:1979ih,
  Collins:1980ih,Sen:1981sd,Korchemsky:1988pn,Collins:1989bt,Magnea:1990zb,Dixon:2008gr}:
\begin{equation}
{\cal A}\sim {\cal H}\cdot{\cal S}\cdot\prod_{i=1}^L\frac{J_i}{{\cal J}_i}.
\label{ampfac}
\end{equation}
Here the left-hand side represents an amplitude with a fixed number $L$ of 
outgoing hard particles, where it is assumed that ultraviolet renormalisation 
has already taken place. The first factor on the right-hand side is the 
{\it hard function}, and is finite in four dimensions (i.e. contains no IR
singularities). We can think of this as a subamplitude for the outgoing hard
particles, where no soft gluons are emitted. The second factor ${\cal S}$ is 
the {\it soft function}, and collects all singularities associated with gluon
emissions whose 4-momenta each tend to zero. The final
factors, one for each outgoing particle, collect {\it collinear 
singularities}, i.e. divergences associated with gluons emitted parallel to 
a given outgoing hard particle, in the case that this is massless. We will not
be concerned with these here, but let us briefly remark that $J_i$
is called a {\it jet function}, and contains collinear singularities associated
with outgoing particle $i$. Because gluons can be soft and collinear at the 
same time, one double counts these singularities by including them both in the
soft function ${\cal S}$ and the jet functions $\{J_i\}$. One corrects for this
double counting by dividing by {\it eikonal jet functions} ${\cal J}_i$, which
can be thought of as the jet functions $J_i$ evaluated in the soft limit. \\

From now on we focus exclusively on the soft function ${\cal S}$ which, as 
described above, collects all singularities arising from multiple gluon soft
gluon emission, including overlapping soft / collinear divergences. We will
see that the soft function has a Feynman diagram intepretation {\it by itself},
so that it will be possible to calculate the soft function independently of
the particular details of a given hard scattering process (other than the
number of hard outgoing particles, and their momenta). Furthermore, we will see
that the soft function ${\cal S}$ has an exponential form, such that one may
schematically write
\begin{equation}
{\cal S}\sim\exp\left[\sum_n\alpha^n\,c_n\right],
\label{softexp}
\end{equation}
where $\alpha$ is the coupling constant, and the $\{c_n\}$ are coefficients
which depend upon the momenta of the hard outgoing particles. This formula
is a highly useful result, for the following reason: because soft gluon 
corrections appear in an exponent, they are summed up to all orders in 
perturbation theory. Expanding the exponential as a Taylor series
in $\alpha$ gives terms to infinite order, where IR divergent
contributions at higher orders are related to singularities at lower orders.
Successive terms in $\alpha$ in the exponent sum up successive towers of
divergent contributions in the perturbation series. We see, then, that 
exponentiation is crucial to the resummation of soft gluon contributions, and
that the problem of classifying IR divergences to all orders in perturbation
theory is equivalent to the study of the exponent of eq.~(\ref{softexp}). \\

We can go further than this. It turns out that, in both abelian and non-abelian
gauge theories (as well as perturbative quantum gravity), the {\it exponent} of
${\cal S}$ can be given a Feynman diagram interpretation directly. That is,
the cofficients $c_n$ can be obtained from a special class of Feynman diagrams.
The particular properties these diagrams have depend on which theory we are
considering. For now, we will refer to any such diagram as a {\it web}:\\

\begin{center}
\fbox{A {\it web} is a diagram that contributes to the exponent of the soft 
function.}\\
\end{center}

We will make this definition more precise (and even revise it) in the
various theories to be examined in what follows, but this first
definition of a web efficiently encapsulates the main idea. Before
examining specific theories, it is useful to motivate why webs are
useful in the first place. There are a number of reasons, which
include the following:
\begin{enumerate}
\item By definition, webs allow one to compute the exponents of soft
gluon amplitudes directly. This is in contrast to the traditional way of 
calculating the exponent at a given order, which is to subtract from the
(unexponentiated) amplitude those terms which arise from the exponentiation
of lower order terms. This involves a large amount of unnecessary work, and
it is much more efficient to be able to work directly in the exponent.
\item Webs are, in an important sense, complete Feynman
  diagrams. Thus, some finite contributions may exponentiate as well
  as parts which are formally singular.  Such finite contributions can
  be numerically large in many scattering processes, so that
  resummation of them is a useful means of increasing the precision of
  theoretical predictions.
\item Recent work has begun to generalise factorisation formulae beyond
  the pure soft gluon approximation (also referred to as the {\it
    eikonal approximation} for the outgoing hard
  particles)~\cite{Dokshitzer:2005bf,Laenen:2008gt,Laenen:2008ux,
    Moch:2009mu,Grunberg:2009yi,Moch:2009hr,Soar:2009yh,Laenen:2010uz,
    Almasy:2010wn,Ball:2013bra,Altinoluk:2014oxa,Apolinario:2014csa,deFlorian:2014vta,
    Presti:2014lqa,Bonocore:2014wua,Bonocore:2015esa,White:2014qia}.
  Webs may be an essential ingredient of this programme in the most
  general case.
\item Webs, and their associated mathematical structures, may give new 
insights into the structure of IR singularities in a variety of theories,
including in particular QCD and ${\cal N}=4$ Super-Yang-Mills theory.
\item There are links between webs in non-abelian gauge theories and
  interesting problems in the field of enumerative combinatorics. Some
  of these problems may have practical applications in computer
  science, thus there is potentially an intriguing possibility that
  lessons from quarks and gluons can be applied to e.g. network
  transfer protocols, or search algorithms.
\item Recent progress hints at systematic methods for computing web
  diagrams~\cite{Gardi:2013saa,Falcioni:2014pka,Laenen:2015jia,Laenen:2014jga,Grozin:2014hna,Grozin:2014axa,Henn:2013wfa,Henn:2012qz,Correa:2012nk,Correa:2012at}. This
  offers the hope of being able to compute all-order properties in the
  {\it exponents} of amplitudes, which constitutes an unprecented look
  at intricate structures in perturbation theory.
\end{enumerate}
Having introduced the basic idea of webs, and also having hopefully conveyed
why they might be useful, let us now begin to examine the simplest case of
webs - namely, the exponentiation of soft photon corrections in abelian gauge
theory. 

\section{Abelian Exponentiation}
\label{sec:abelian}
In this section we consider QED, and show in a specific example that soft
photon contributions exponentiate, where the exponent can be interpreted 
directly in terms of a type of Feynman diagram. This analysis is similar to
that originally carried out in~\cite{Yennie:1961ad}. \\

\begin{figure}
\begin{center}
\scalebox{0.8}{\includegraphics{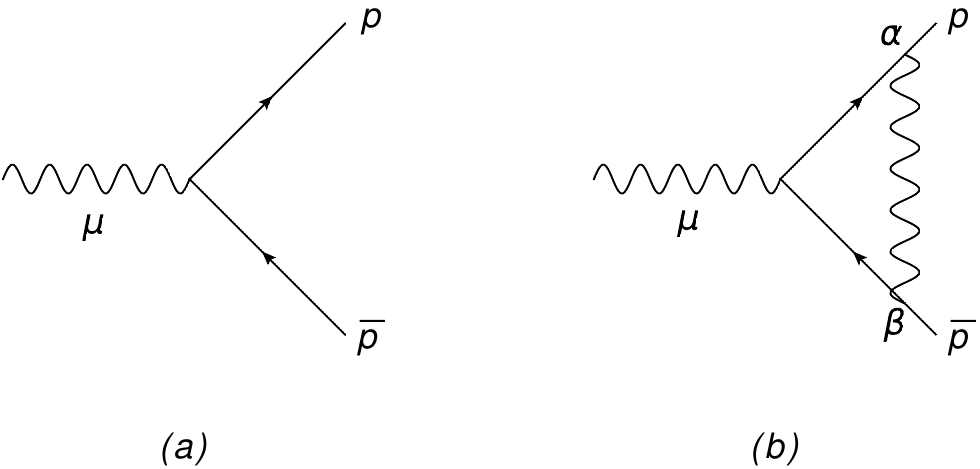}}
\caption{Amplitude for an offshell photon decaying to a fermion-antifermion
pair: (a) leading order; (b) dressed with a single photon emission.}
\label{fig:amp}
\end{center}
\end{figure}

\begin{figure}
\begin{center}
\scalebox{0.8}{\includegraphics{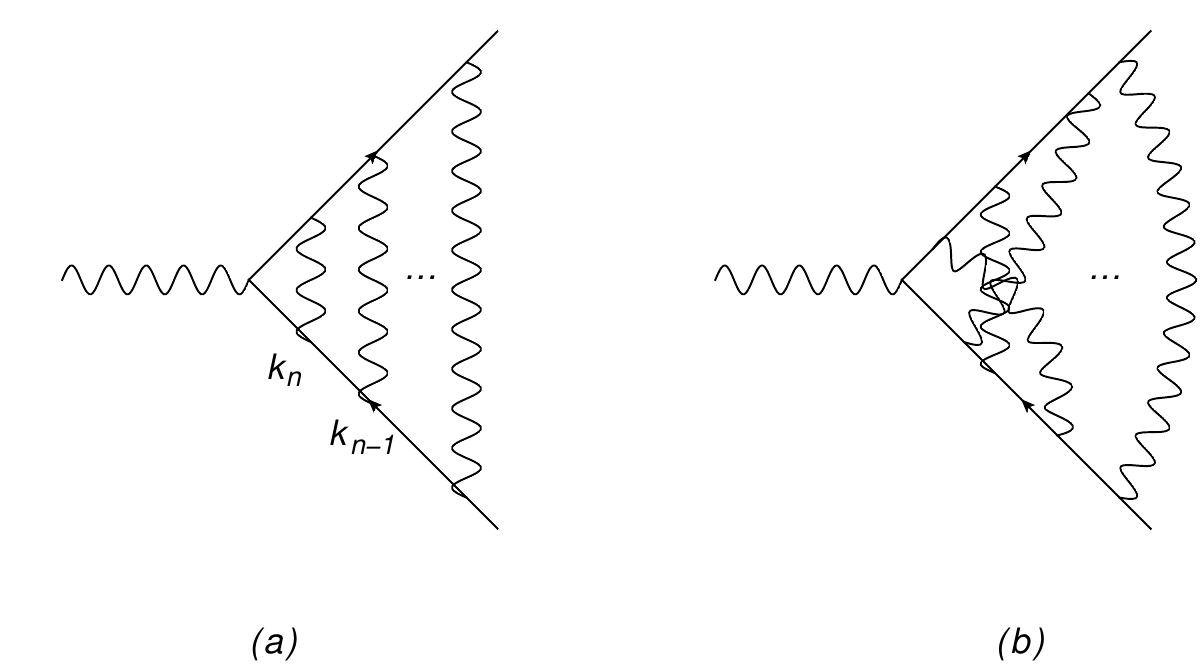}}
\caption{Two diagrams which contribute at ${\cal O}(\alpha^n)$, where
in the left-hand diagram we label the momentum of the $i^{\rm th}$ gluon
by $k_i$.}
\label{fig:nphoton}
\end{center}
\end{figure}

We begin by considering the simple scattering amplitude shown 
in~\ref{fig:amp}(a), consisting of an (off-shell) photon decaying into a 
massless fermion / anti-fermion pair. We label the 4-momenta of the latter by 
$p$ and $\bar{p}$ respectively, as shown in the figure. Ignoring coupling 
factors for brevity, the amplitude in momentum space is given by
\begin{equation}
{\cal A}_0=\bar{u}(p)\gamma^\mu v(\bar{p}),
\label{A0def}
\end{equation}
where $\mu$ is the Lorentz index of the offshell photon. 
Let us now consider adding a single photon emission, as shown in 
figure~\ref{fig:amp}(b). The corresponding amplitude is
\begin{equation}
{\cal A}_1=\bar{u}(p)\gamma^\alpha\frac{(\slsh{p}-\slsh{k})}{(p-k)^2}
\gamma^\mu\frac{(\slsh{k}-\slsh{\bar{p}})}{(k-\bar{p})^2}\gamma^\beta 
v(\bar{p})D_{\alpha\beta}(k),
\label{A1def}
\end{equation}
where $D_{\alpha\beta}(k)$ denotes the photon propagator, and we do not show
explicitly the integral over the gluon momentum $k$. One may now consider the
soft limit $k\rightarrow 0$. In this limit, we can ignore the factors of 
$\slsh{k}$ in the numerator of eq.~(\ref{A1def}), and neglect the $k^2$ terms
in the denominators. Using also the fact that $p^2=0$, one finds
\begin{equation}
{\cal A}_1\rightarrow\bar{u}(p)\gamma^\alpha\frac{\slsh{p}}{-2p\cdot k}
\gamma^\mu\frac{\slsh{\bar{p}}}{2\bar{p}\cdot k}\gamma^\beta 
v(\bar{p})D_{\alpha\beta}(k),
\label{A1def2}
\end{equation}
We can simplify this expression by using the anticommutation property of the
Dirac matrices, $\{\gamma^\mu,\gamma^\nu\}=2g^{\mu\nu}$, to write
\begin{align}
\bar{u}(p)\gamma^\alpha\slsh{p}&=-\bar{u}(p)\slsh{p}\gamma^\alpha+2p^\alpha
\bar{u}(p)\label{dirac1}\\
\slsh{\bar{p}}\gamma^\beta v(\bar{p})&=-\gamma^\beta\slsh{\bar{p}} v(\bar{p})
+2\bar{p}^\beta v(\bar{p}).\label{dirac2}
\end{align}
The Dirac equations for the fermion and antifermion spinors 
($\bar{u}(p)\slsh{p}=\slsh{\bar{p}}v(p)=0$) imply that the first terms of 
eqs.~(\ref{dirac1}) and~(\ref{dirac2}) vanish. Then eq.~(\ref{A1def2}) becomes
\begin{equation}
{\cal A}_1\rightarrow\left[\bar{u}(p)\gamma^\mu v(\bar{p})\right]
\left(-\frac{p^\alpha}{p\cdot k}\right)
\left(\frac{\bar{p}^\alpha}{\bar{p}\cdot k}\right)D_{\alpha\beta}(k).
\label{A1def3}
\end{equation}
We may note the following from this expression:
\begin{itemize}
\item The tree-level interaction ${\cal A}_0$ factors out, and is not IR
divergent. Thus, this is an explicit example of the factorisation of soft 
physics from hard, which is embodied more generally by eq.~(\ref{ampfac}).
The essential reason for this factorisation is that a soft photon has an 
infinite Compton wavelength, and therefore cannot resolve the details of the
hard interaction (which, by the uncertainty principle, takes place over 
relatively short distance scales).
\item The emission of a soft photon from the fermion or anti-fermion line is
described by the {\it eikonal Feynman rule}
\begin{equation}
\frac{p^\mu}{p\cdot k},
\label{eikrule}
\end{equation}
where the sign is different on the two lines due to the fact that the momentum
$k$ is outgoing at one vertex, but incoming at the other. Here we have 
considered soft photon emission from fermion lines, but in fact the same 
Feynman rule would have resulted had we considered scalar or vector particles. 
The reason for this is again due to the large Compton wavelength of the soft 
photon, which cannot resolve the magnetic moment of the emitting particle, 
hence is insensitive to the spin. 
\item We can think of the factors multiplying the tree-level interaction
${\cal A}_0$ as a Feynman diagram by itself, albeit with effective Feynman 
rules. This is illustrated pictorially in figure~\ref{fig:oneloopfac},
\begin{figure}
\begin{center}
\scalebox{0.8}{\includegraphics{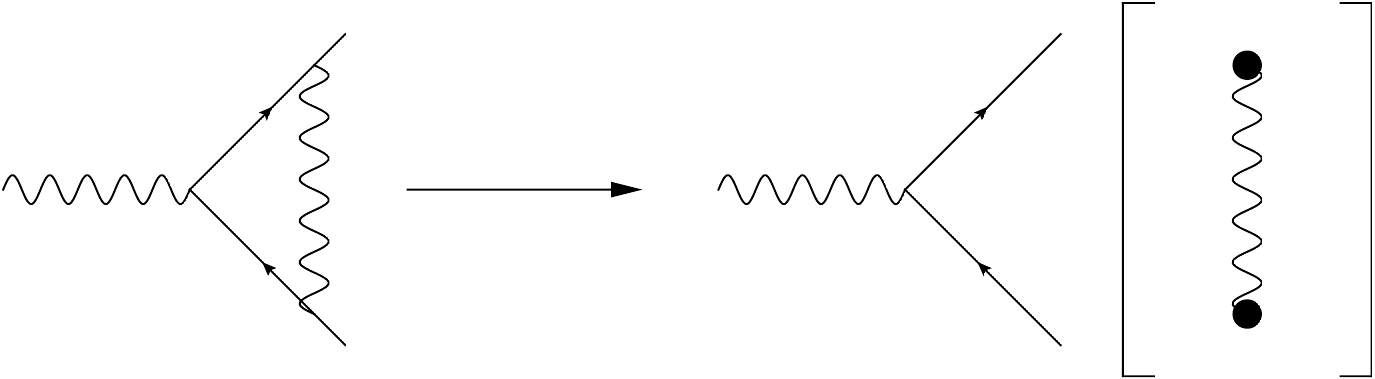}}
\caption{Pictorial depiction of the factorisation of the one loop graph
of figure~\ref{fig:amp}(b) in the soft photon limit, where $\bullet$ denotes
an eikonal Feynman rule.}
\label{fig:oneloopfac}
\end{center}
\end{figure}
where we draw the one loop amplitude in the soft limit as the product of the
tree level diagram with an effective Feynman diagram involving two eikonal
Feynman rules joined by a photon propagator. The latter part is a 
{\it subdiagram} of the full amplitude which spans the hard external lines,
and generates the IR singularities of the amplitude following the integration
over the photon momentum $k$. 
\end{itemize}
Having observed the factorisation of soft photon physics at the one loop level,
the next step is to consider what happens when we add any number of soft photon
emissions. At a given order ${\cal O}(\alpha^n)$ there are many different 
possibilities, two of which are shown in figure~\ref{fig:nphoton}(a). 
Each diagram gives a somewhat complicated expression, as we will see in the
following. However, a very simple result is obtained in the soft limit for the
sum of all diagrams at ${\cal O}(\alpha^n)$, which is illustrated pictorially
in figure~\ref{fig:nphotonsum}. 
\begin{figure}
\begin{center}
\scalebox{0.8}{\includegraphics{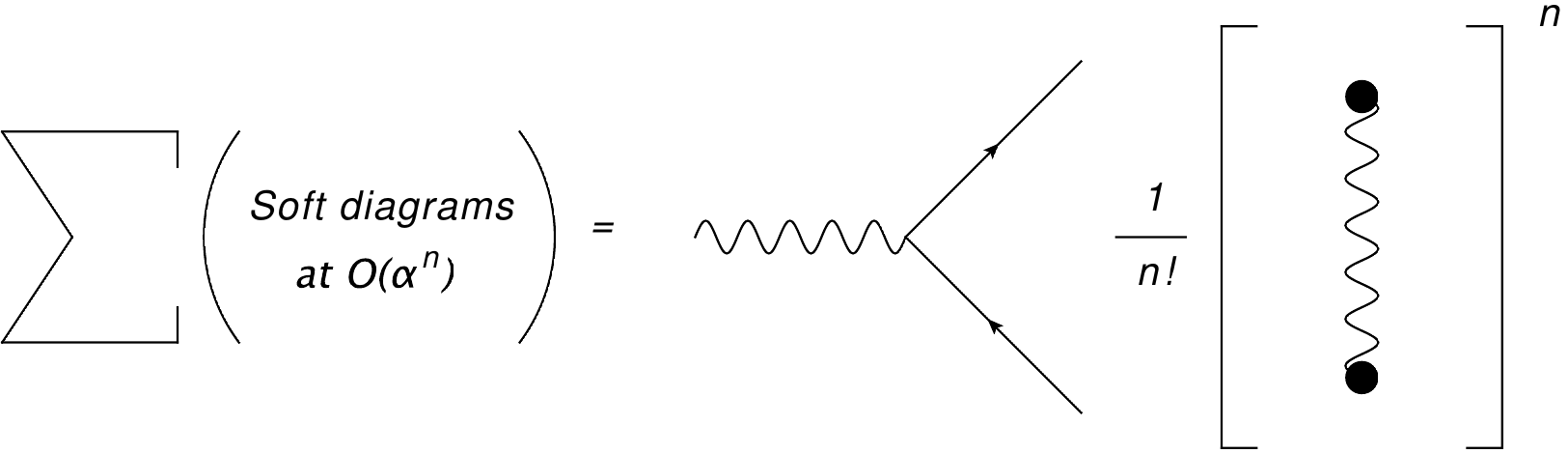}}
\caption{Pictorial representation of the sum of all soft photon diagrams at
${\cal O}(\alpha^n)$, using the same notation as figure~\ref{fig:oneloopfac}.}
\label{fig:nphotonsum}
\end{center}
\end{figure}
This shows that in the sum of all soft photon diagrams, the tree-level
interaction ${\cal A}_0$ again factors out. The remainder is simply
related to the one loop diagram raised to the power $n$, where the
product is meant in the same sense as in
figure~\ref{fig:oneloopfac}. We may prove this result as follows.\\

Let us first consider the particular diagram shown in 
figure~\ref{fig:nphoton}(a), consisting of $n$ uncrossed photon exchanges.
This is usually referred to as a {\it ladder diagram} in the literature. 
The expression for this diagram has the general form
\begin{equation}
{\cal A}^{(n)}_{a}=\bar{u}(p)Q^{\alpha_1\ldots\alpha_n}\,\gamma^\mu\,
\bar{Q}^{\beta_1\ldots\beta_n}D_{\alpha_1\beta_1}(k_1)\ldots 
D_{\alpha_n\beta_n}(k_n),
\label{Ana1}
\end{equation}
where the tensors $Q^{\alpha_1\ldots\alpha^n}$ and
$\bar{Q}^{\beta_1\ldots \beta_n}$ collect all factors from the fermion
and anti-fermion lines respectively, and we have included the product
of photon propagators such that the $i^{\rm th}$ photon has Lorentz
indices $\alpha_i$, $\beta_i$. Again, we do not explicitly show the
integrals over the loop momenta $k_i$, and ignore coupling constants
etc. The factors associated with the quark line are
\begin{equation}
\bar{u}(p)Q^{\alpha_1\ldots\alpha_n}=\frac{\bar{u}(p)\gamma^{\alpha_1}
(\slsh{p}-\slsh{k}_1)\gamma^{\alpha_2}(\slsh{p}-\slsh{k}_1-\slsh{k}_2)
\ldots \gamma^{\alpha_n}(\slsh{p}-\slsh{k_1}-\ldots-\slsh{k}_n)}
{(p-k_1)^2(p-k_1-k_2)^2\ldots(p-k_1-k_2-\ldots-k_n)^2},
\label{uQ1}
\end{equation}
which in the soft limit of $k_i\rightarrow 0$ for each $i$ becomes
\begin{equation}
\bar{u}(p)Q^{\alpha_1\ldots\alpha_n}=\frac{\bar{u}(p)\gamma^{\alpha_1}\slsh{p}
\gamma^{\alpha_2}\slsh{p}\ldots\gamma^{\alpha_n}\slsh{p}}
{[-2p\cdot k_1][-2p\cdot(k_1+k_2)][-2p\cdot(k_1+k_2+\ldots+k_n)]}.
\label{uQ2}
\end{equation}
As in the case of one photon emission, one may anticommute the factors of 
$\slsh{p}$ through the Dirac matrices, to act on the spinor $\bar{u}(p)$. 
All terms involving $\bar{u}(p)\slsh{p}$ vanish from the Dirac equation,
leaving
\begin{equation}
\bar{u}(p)Q^{\alpha_1\ldots\alpha_n}=\frac{\bar{u}(p)(-1)^np^\alpha_1\ldots
p^{\alpha_n}}{p\cdot k_1\,p\cdot(k_1+k_2)\,p\cdot(k_1+\ldots+k_n)}.
\label{uQ3}
\end{equation}
Similarly for the factors on the anti-fermion line one finds 
\begin{equation}
\bar{Q}^{\beta_1\ldots\beta_n}v(\bar{p})=\frac{\bar{p}^{\beta_1}\ldots
\bar{p}^{\beta_n}}{\bar{p}\cdot k_1\,\bar{p}\cdot(k_1+k_2)\ldots
\bar{p}\cdot(k_1+\ldots+k_n)}v(p),
\label{Qbar1}
\end{equation}
such that the amplitude of eq.~(\ref{Ana1}) becomes
\begin{align}
{\cal A}^{(n)}_a&={\cal A}_0\Big\{\frac{(-1)^np^{\alpha_1}\ldots p^{\alpha_n}
\,\bar{p}^{\beta_1}\ldots\bar{p}^{\beta_n}}{p\cdot k_1\,p\cdot(k_1+k_2)
\ldots p\cdot(k_1+\ldots+k_n)\,\bar{p}\cdot k_1\,\bar{p}\cdot(k_1+k_2)
\ldots \bar{p}\cdot(k_1+\ldots+k_n)}\notag\\
&\quad\times D_{\alpha_1\beta_1}(k_1)\ldots D_{\alpha_n\beta_n}(k_n)\Big\}.
\label{Ana2}
\end{align}
Again we see that the tree-level interaction ${\cal A}_0$ has factored out,
and clearly this will happen for any such soft photon diagram at arbitrary 
order. Note that the denominators in eq.~(\ref{Ana2}) indicate a complicated
dependence on the photon momenta $k_i$. These are coupled nontrivially along 
both the fermion and the anti-fermion line, so a given photon somehow knows
about all other photons lying between itself and the final state hard 
particles. However, we have only considered one diagram at this order, and must
now consider summing over all such diagrams.\\

\begin{figure}
\begin{center}
\scalebox{0.8}{\includegraphics{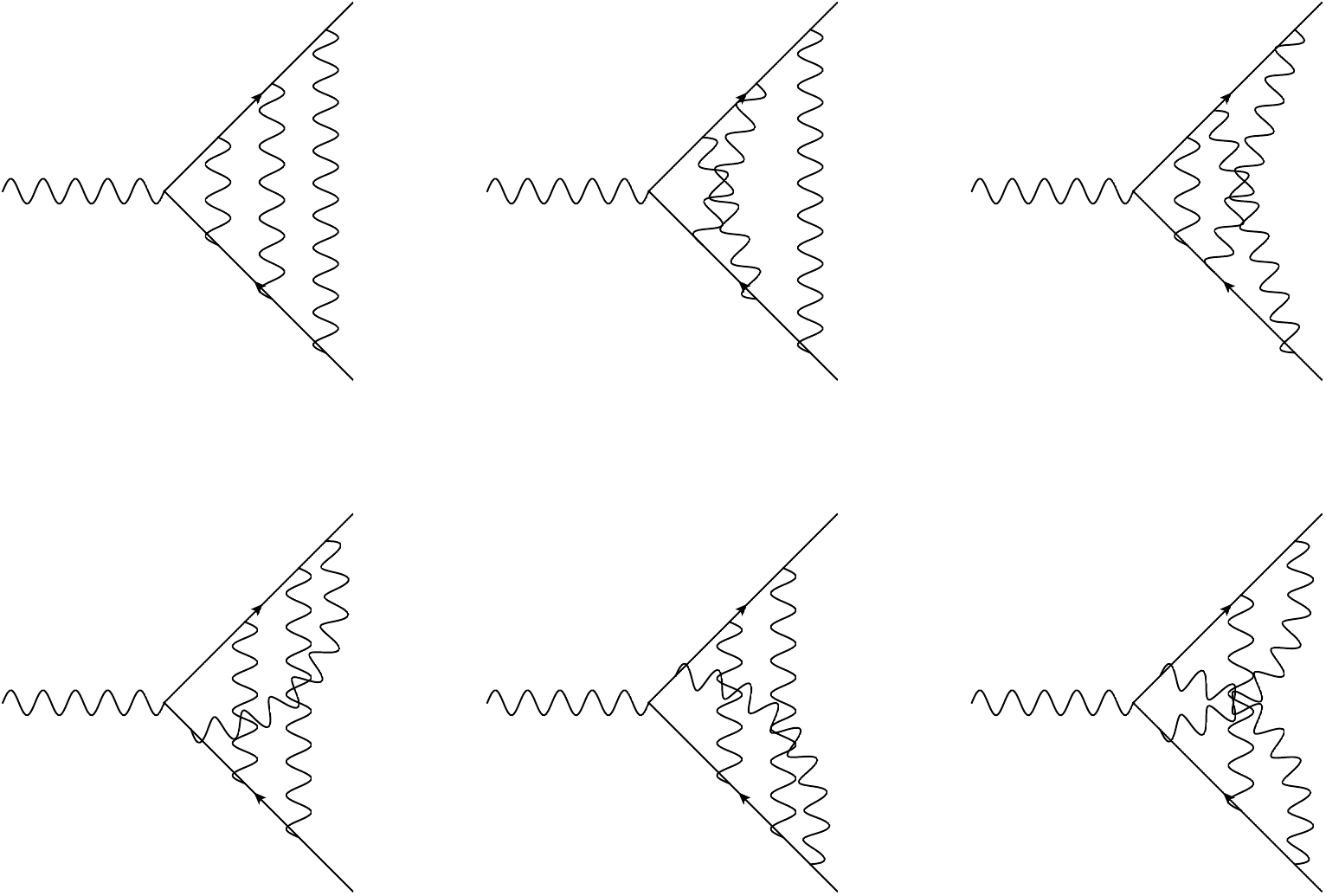}}
\caption{Complete set of diagrams involving three photon emissions at
${\cal O}(\alpha^3)$.}
\label{fig:n3sum}
\end{center}
\end{figure}

This set of diagrams is illustrated for the $n=3$ case in 
figure~\ref{fig:n3sum}, where we see that there are six diagrams involving 
three photon emissions.
We can think of obtaining this set by fixing the order of the photon emissions
on the lower (anti-fermion) line, and summing over all permutations of 
emission orders on the upper (fermion) line. Thus, the total number of diagrams
should be the same as the total number of permutations of three objects, which
is $3!=6$ as observed. Returning to the general case of $n$ photon emissions, 
we may write the complete sum over all diagrams from eq.~(\ref{Ana2}) as
\begin{align}
{\cal A}^{(n)}_a&={\cal A}_0\Big\{\frac{(-1)^np^{\alpha_1}\ldots p^{\alpha_n}
\,\bar{p}^{\beta_1}\ldots\bar{p}^{\beta_n}}{\bar{p}\cdot k_1\,\bar{p}
\cdot(k_1+k_2)\ldots \bar{p}\cdot(k_1+\ldots+k_n)}D_{\alpha_1\beta_1}
(k_1)\ldots D_{\alpha_n\beta_n}(k_n)\notag\\
&\quad\times\sum_\pi\frac{1}{p\cdot k_{\pi_1}\,p\cdot(k_{\pi_1}+k_{\pi_2})
\ldots p\cdot(k_{\pi_1}+\ldots+k_{\pi_n})}\Big\}.
\label{Ana3}
\end{align}
Here $\pi$ labels a permutation of $(1,2,\ldots,n)$, such that this maps to
$(\pi_1,\pi_2,\ldots,\pi_n)$. The rather formidable looking formula of 
eq.~(\ref{Ana3}) can be drastically simplified if we make use of the 
{\it eikonal identity}
\begin{equation}
\sum_\pi\frac{1}{p\cdot k_{\pi_1}\,p\cdot(k_{\pi_1}+k_{\pi_2})
\ldots p\cdot(k_{\pi_1}+\ldots+k_{\pi_n})}=\prod_i\frac{1}{p\cdot k_i}.
\label{eikid}
\end{equation}
In order to make sense of this result, it is instructive to consider the
case of $n=2$, in which eq.~(\ref{eikid}) reduces to
\begin{equation}
\frac{1}{p\cdot k_1\,p\cdot(k_1+k_2)}+\frac{1}{p\cdot k_2\,p\cdot(k_1+k_2)}
=\frac{p\cdot k_1+p\cdot k_2}{p\cdot k_1\,p\cdot k_2\,p\cdot(k_1+k_2)}
=\frac{1}{p\cdot k_1\,p\cdot k_2},
\label{eikid2}
\end{equation}
which is indeed the stated result. We see that the eikonal identity is a sort
of partial fraction identity, and the general result could be proven e.g. by
induction from lower orders. \\

\begin{center}
\parbox{5in}{{\bf Problem 1}. Show by explicit calculation that the eikonal
identity works for the $n=3$ case.}
\end{center}

Substituting eq.~(\ref{eikid}) into eq.~(\ref{Ana3}), the latter becomes
\begin{align}
{\cal A}^{(n)}_a&={\cal A}_0\Big\{\frac{(-1)^np^{\alpha_1}\ldots p^{\alpha_n}
\,\bar{p}^{\beta_1}\ldots\bar{p}^{\beta_n}}{\bar{p}\cdot k_1\,\bar{p}
\cdot(k_1+k_2)\ldots \bar{p}\cdot(k_1+\ldots+k_n)}D_{\alpha_1\beta_1}
(k_1)\ldots D_{\alpha_n\beta_n}(k_n)\notag\\
&\quad\times\prod_{i=1}^n\frac{1}{p\cdot k_i}\Big\}.
\label{Ana4}
\end{align}
This result is already a lot simpler than eq.~(\ref{Ana3}) for one
important reason: the photons on the upper line are now completely
decoupled from each other. This is not true for the photons on the
lower line. However, one may achieve this by exploiting the fact that
all of the momenta $k_i$ are dummy variables which are ultimately
integrated over. Furthermore, the product of photon propagators
contracted with the momentum factors is symmetric under interchange of
any two momenta, subject to suitable relabelling of dummy Lorentz
indices. Thus, by relabelling the integration variables one may
replace the particular combination of photon momenta appearing in the
first line of eq.~(\ref{Ana4}) with any permutation of these momenta.
That is, one may write
\begin{align}
&\int d^dk_1\ldots\int d^dk_n\frac{1}{\bar{p}\cdot k_1\,\bar{p}
\cdot(k_1+k_2)\ldots \bar{p}\cdot(k_1+\ldots+k_n)}\notag\\
&=\frac{1}{n!}\int d^dk_1\ldots\int d^dk_n\sum_\pi\frac{1}{\bar{p}\cdot 
k_{\pi_1}\,\bar{p}\cdot(k_{\pi_1}+k_{\pi_2})\ldots \bar{p}\cdot
(k_{\pi_1}+\ldots+k_{\pi_n})}.
\label{intks}
\end{align}
In the second line, we have used the fact that each term gives the same result
(by relabelling dummy integration variables), and that there are $n!$ such 
permutations. Substituting this result into eq.~(\ref{Ana4}) and 
tidying things up (again omitting the integrations over the momenta for 
brevity), one ultimately finds
\begin{align}
{\cal A}^{(n)}&={\cal A}_0\frac{1}{n!}\prod_{i=1}^n\left(\frac{-p^{\alpha_i}}
{p\cdot k_i}\right)\left(\frac{\bar{p}^{\beta_i}}{\bar{p}\cdot k_i}\right)
D_{\alpha_i\beta_i}(k_i)\notag\\
&={\cal A}_0\frac{1}{n!}\left[\left(\frac{-p^{\alpha}}
{p\cdot k}\right)\left(\frac{\bar{p}^{\beta}}{\bar{p}\cdot k}
\right)D_{\alpha\beta}(k)\right]^n,
\label{Ana5}
\end{align}
which is precisely the result expressed pictorially in 
figure~\ref{fig:nphotonsum}, namely that the sum over all $n$ photon graphs
in the soft limit is simply related to the $n^{\rm th}$ power of the one loop
graph. \\

\begin{figure}
\begin{center}
\scalebox{0.6}{\includegraphics{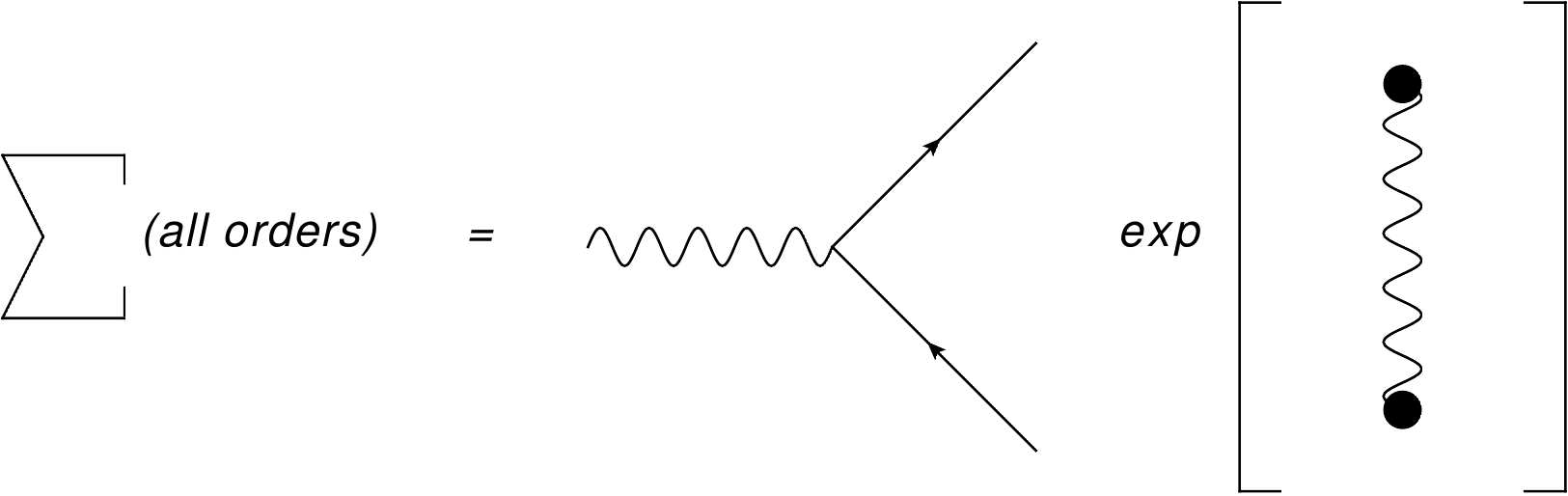}}
\caption{Pictorial representation of the exponentiation result of 
eq.~(\ref{expres}).}
\label{fig:exp1}
\end{center}
\end{figure}

\begin{figure}
\begin{center}
\scalebox{1.0}{\includegraphics{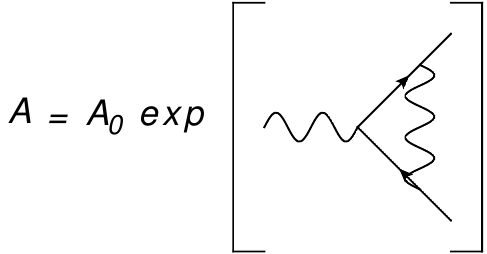}}
\caption{Alternative drawing of figure~\ref{fig:exp1}, making clear
which outgoing particles emit the soft photon.}
\label{fig:exp2}
\end{center}
\end{figure}

The astute reader will have noticed that eq.~(\ref{Ana5}) (which includes the
leading order result) is the $n^{\rm th}$ term in the Taylor expansion of an 
exponential. Thus, one may explicitly compute the sum of all diagrams involving
any number of single soft photon emissions, to all orders in perturbation 
theory! The result is
\begin{equation}
{\cal A}={\cal A}_0\exp\left[\int\frac{d^dk}{(2\pi)^d}\left(\frac{-p^{\alpha}}
{p\cdot k}\right)\left(\frac{\bar{p}^{\beta}}{\bar{p}\cdot k}
\right)D_{\alpha\beta}(k)\right],
\label{expres}
\end{equation}
where we have reinstated the integral over the gluon momentum. This result 
is shown using our pictorial notation in figure~\ref{fig:exp1}. 
In later sections, we will end up considering diagrams with many hard
particles connected by soft gauge bosons. Thus, it is useful to redraw
figure~\ref{fig:exp1} so as to include the hard interaction in the
exponent, as shown in figure~\ref{fig:exp2}. The purpose of this is
simply to clearly display which lines the photon is emitted between -
we {\it do not mean} that the hard interaction itself
exponentiates. \\

\begin{figure}
\begin{center}
\scalebox{1.0}{\includegraphics{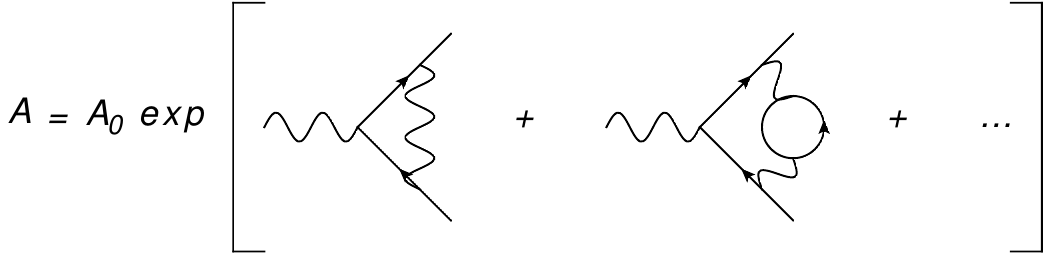}}
\caption{Complete result for abelian exponentiation, in which all possible
connected soft photon subgraphs appear in the exponent.}
\label{fig:exp3}
\end{center}
\end{figure}

We have now derived our first example of exponentiation. Some further comments:
\begin{enumerate}
\item[(i)] The one loop graph in the exponent of eq.~(\ref{expres}) has an
IR divergence. Were we to explicitly expand the exponential, we would thus 
find that IR singularities at higher orders in the amplitude are related to
singularities at lower orders.
\item[(ii)] Here we considered only graphs with single photon
  emissions, finding that these are entirely given by the exponential
  of the one loop graph.  In fact, this is not the whole story - soft
  photons could also connect off the external lines, via fermion
  bubbles. The complete result is then that the exponent contains all
  connected soft photon subgraphs, as depicted in
  figure~\ref{fig:exp3}. In principle one can prove this using a
  generalisation of the above method~\cite{Yennie:1961ad}. We will
  instead prove this result using alternative means in the following
  section.
\item[(iii)] Above, we only considered two outgoing hard particles. However,
the result can be extended to any number of external lines, where again the
exponent contains all possible connected subgraphs which span the hard particle
lines. 
\item[(iv)] Things are, unsurprisingly, more complicated in non-abelian gauge 
theories. Crucial to the above argument was that we could sum over all 
permutations of photon emissions, where these were weighted equally. This is
no longer the case in e.g. QCD, where the emission vertices carry colour 
matrices ${\bf T}^a$ which do not commute. Thus, an additional level of
combinatoric complexity arises.
\item[(v)] Things are not much more complicated in perturbative
  quantum gravity. This is a theory that shares features of QED and
  QCD. It is QED-like in that soft graviton emission vertices commute
  with each other (i.e. do not carry a non-commuting matrix-valued
  charge). It is QCD-like in that there are three and four graviton
  vertices, which mimic the corresponding gluon self-interactions in
  QCD. However, the charge in gravity is the 4-momentum of a
  particle. Thus, all multiple vertices between soft (zero momentum)
  gravitons vanish. A careful analysis also shows that diagrams in
  which gravitons couple to matter loops also
  vanish~\cite{Naculich:2011ry,Akhoury:2011kq}. Hence, IR divergences
  are given to all orders in quantum gravity purely from the
  exponentiation of the one-loop result, a property known as {\it
    one-loop exactness}. In other words, the structure of IR
  divergences in quantum gravity is, remarkably, simpler than in QED!
\end{enumerate}

In section~\ref{sec:intro}, we defined a web as being a diagram which 
contributes to the exponent of the soft function. We have now seen our first
examples of webs, and the results of this section can be compactly summarised
as follows:\\

\begin{center}
\fbox{Webs in abelian gauge theories are connected subdiagrams.}\\
\end{center}

At this point, we could leave the case of QED in favour of exploring
non-abelian gauge theories. The traditional treatment of webs in the
latter case proceeds by generalising the eikonal identity to take into
account non-trivial colour structure. One may then explicitly solve
for the structure of webs in scattering amplitudes which involve only
two coloured external
particles~\cite{Gatheral:1983cz,Frenkel:1984pz,Sterman:1981jc}~\footnote{For
  a pedagogical exposition, see~\cite{Berger:2003zh}.}. We choose not
to follow such an approach here. Firstly, there is the lack of
generality implied by the restriction to processes involving only two
coloured particles (although three particles can also be treated
straightforwardly using similar
methods~\cite{Berger:2003zh})~\footnote{See
  also~\cite{Mitov:2010rp,Vladimirov:2014wga,Vladimirov:2015fea} for
  an examination of the general multileg case.}. Secondly, one may
develop a more general approach for deriving exponentiation, which
works for both abelian and non-abelian gauge theories, and any number
of outgoing particles. In order to introduce this method, we will
start by rederiving the results of abelian exponentiation. This is the
subject of the following section.

\section{The Path Integral Approach}
\label{sec:path}
\begin{figure}
\begin{center}
\scalebox{1.0}{\includegraphics{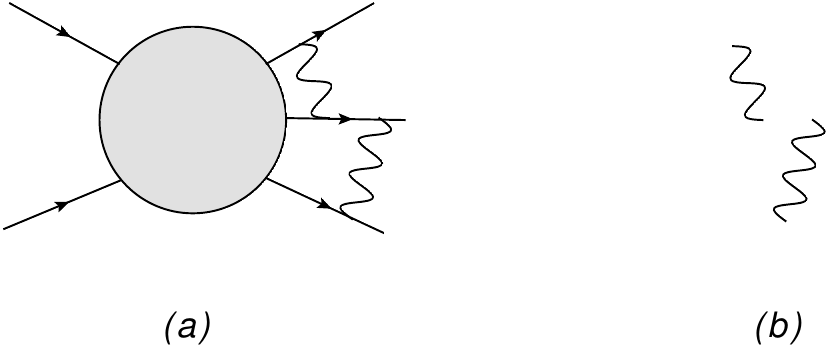}}
\caption{(a) An example scattering amplitude involving five external hard 
particles, and two soft photons; (b) soft photon subdiagram.}
\label{fig:subdiag}
\end{center}
\end{figure}

In the previous section, we saw that the soft function itself has a Feynman
diagram interpretation, where these diagrams consisted of effective Feynman
rules for emission of soft gluons from hard outgoing particles. The 
propagators joining these vertices were the usual propagators arising from 
the action for the soft gauge field. Furthermore, these diagrams were
{\it subdiagrams} in the full amplitude. That is, if one removes the hard 
interaction and the hard outgoing particle lines, what is left is the soft
gluon diagram. This is illustrated in figure~\ref{fig:subdiag}, for a 
scattering amplitude involving five external particle legs. 
In practice, as we saw at the end of the previous section, it is useful to
also draw the external lines in the soft photon subdiagram, so as to indicate
which external line a given soft photon was emitted from. This matters given 
that the hard momentum $p^\mu$ enters the eikonal Feynman rule.\\

Given that Feynman diagrams exist purely for the soft subdiagrams, it
follows that there must be a quantum field theory for the soft gauge
field by itself, whose diagrams are the soft photon subdiagrams. One
way of specifying a QFT is to write down its {\it generating
  functional}, from which all Feynman diagrams can be derived. Here we
will simply write down the generating functional which generates soft
photon subdiagrams (see~\cite{Laenen:2008gt} for a full
derivation). It is given by~\footnote{Here we have absorbed a factor
  of the coupling constant into the gauge field.}
\begin{equation}
Z=\int{\cal D}A_\mu e^{iS[A_\mu]}\prod_{k=1}^Le^{i\int dx_k^\mu A_\mu(x_k)},
\label{genfunc}
\end{equation}
where $A_\mu$ is the (soft) gauge field with action $S[A_\mu]$, and
the product is over external particles, $L$ in total.  The exponential
factors contain the space-time trajectory of the $k^{\rm th}$
particle, which is given by
\begin{equation}
x_k^\mu=n_k^\mu t,
\label{xkdef}
\end{equation}
where $n_k^\mu$ is a (constant) 4-vector proportional to the 4-momentum of the
particle, $p_k^\mu$. Note that we have taken each outgoing particle as having
a zero initial position. Thus, all the hard particles originate from the 
origin, which corresponds to having shrunk the hard interaction which produces
these particles to an infinitely small region of space-time. This is an 
approximation of course, but it turns out that corrections to this 
approximation are formally subleading, in that they involve additional powers
of the emitted gluon momenta, which in the soft limit have been taken to zero. 
The physics of this approximation is that soft gluons, as discussed above, have
an infinite Compton wavelength. Thus, they are not able to resolve the 
internal details of the hard interaction, which consequently appears 
pointlike. \\

Let us discuss the form of the generating functional of eq.~(\ref{genfunc}).
Firstly, there is a path integral over all possible configurations of the 
soft gauge field, where each is weighted by a phase factor involving the 
classical action. This is what one expects for the path integral representation
of a quantum field theory. The additional factors (one for each external leg)
are exponential, where the exponent is linear in the gauge field. This thus
acts as a {\it source term} for the soft gauge field. In other words, vertices
will be generated, which enable a soft photon to pop out of the vacuum. 
The form of the exponent tells us that each vertex is located in space-time
along one of the external leg trajectories $x_k^\mu$. We can see what the 
momentum space Feynman rule is for such a vertex by first rewriting the
exponent of each external line factor as
\begin{equation}
\exp\left[i\int dx_k^\mu A_\mu\right]=\exp\left[i\int_0^\infty dt n_k^\mu 
A_\mu(x^\mu)\right],
\label{exfacexp}
\end{equation}
where we have substituted the trajectory of eq.~(\ref{xkdef}). Next, we may
write the soft gauge field in terms of its Fourier transform using
\begin{equation}
A_\mu(x)=\int \frac{d^dk}{(2\pi)^d}\tilde{A}_\mu(k)e^{ik\cdot x}.
\label{Atildedef}
\end{equation}
Substituting this into eq.~(\ref{exfacexp2}), the exponent becomes
\begin{align}
i\int_0^\infty dt n_k^\mu A_\mu(n_k^\mu t)&=\int\frac{d^dk}{(2\pi)^d}
\tilde{A}_\mu(k)i\int_0^\infty dt\, n_k^\mu\, e^{i(k\cdot n_k)t}\notag\\
&=\int\frac{d^dk}{(2\pi)^d}\tilde{A}_\mu(k)\,i\,n_k^\mu\left[
\frac{e^{i(k\cdot n_k)t}}{ik\cdot n_k}\right]^\infty_0\notag\\
&=\int\frac{d^dk}{(2\pi)^d}\tilde{A}_\mu(k)\left(-\frac{n_k^\mu}{n_k\cdot k}
\right).
\label{exfacexp2}
\end{align}
Here we have been rather cavalier in dealing with the upper limit in
the second line. A more careful treatment - including the Feynman
$i\varepsilon$ prescription - gives the same result. By the usual
procedure for reading Feynman rules from the exponent inside the path
integral, the integrand of the momentum integral (represented by the
term in brackets in the last line of eq.~(\ref{exfacexp2})) is the
momentum space Feynman rule. Given that this is evidently invariant
under rescalings of the vector $n_k^\mu\propto p_k^\mu$, we may
replace this Feynman rule with~\footnote{The rescaling symmetry of the
  eikonal Feynman rules is actually broken when the Wilson lines are
  null - see e.g. ~\cite{Gardi:2009zv} for a discussion. For our
  purposes, we can take the Wilson lines to be massive
  i.e. $p_k^2\neq0$.}
\begin{displaymath}
-\frac{p_k^\mu}{p_k\cdot k},
\end{displaymath}
which is the correct Feynman rule for the emission of a soft photon from a hard
external particle, as given in eq.~(\ref{eikrule}). \\

\begin{center}
\parbox{5in}{{\bf Problem 2}. For hard particles emitting soft
  gravitons (described by a spin two field $h_{\mu\nu}$(x)), the
  analogue of the external line factor of eq.~(\ref{exfacexp})
  is~\footnote{This operator was introduced in
    ref.~\cite{Naculich:2011ry}, and further studied in
    refs.~\cite{White:2011yy,Miller:2012an,Melville:2013qca}.}
\begin{displaymath}
\exp\left[-i\frac{\kappa}{2}\int_0^\infty ds \,p^\mu \,p^\nu\,h_{\mu\nu}(sp)
\right],
\end{displaymath}
where $p^\mu$ is the momentum of the particle emitting the gravitons,
$s$ is a parameter such that $x^\mu=sp^\mu$, and $\kappa=\sqrt{16\pi G_N}$,
where $G_N$ is Newton's constant. Show that this leads to an effective 
momentum-space Feynman rule for graviton emission given by
\begin{displaymath}
\frac{\kappa}{2}\frac{p^\mu\,p^\nu}{p\cdot k}.
\end{displaymath}
}
\end{center}

The above discussion justifies why eq.~(\ref{genfunc}) is the correct 
generating functional for soft photon subdiagrams. It generates the right 
Feynman rules for emission of soft photons from external lines. Furthermore,
these will be joined by propagators, as generated by the action for the soft
gauge field $S[A^\mu]$~\footnote{Note that we have been somewhat lapse with
our notation for this action, which also includes couplings to matter
particles, so that fermion loops are generated in soft photon diagrams.}.
Thus, the diagrams generated by eq.~(\ref{genfunc}) are indeed the soft
photon diagrams we are after, which are in turn subdiagrams in the full
amplitude. \\

There is another way to look at the external line factors in 
eq.~(\ref{genfunc}), which is to notice that they are {\it Wilson line}
operators. The general definition of such an operator is~\footnote{As in the
preceding discussion, we have here omitted a factor of the charge $e$ of the 
emitting particle.}
\begin{equation}
\Phi({\cal C})=\exp\left[i\int_{\cal C} dx^\mu \,A_\mu(x)\right],
\label{Wilsondef}
\end{equation}
where ${\cal C}$ is some contour in space-time. Such operators transport gauge
information from point to point in space-time, along the chosen path. What 
eq.~(\ref{genfunc}) then generates is a vacuum expectation value of Wilson 
line operators, evaluated along the classical trajectories of the outgoing 
hard particles. This is in fact a restatement of an old 
result~\cite{Korchemsky:1992xv,Korchemsky:1993uz}, namely that IR 
singularities due to soft gluon emission from hard quarks or gluons can be 
described by dressing the hard particles by Wilson line operators. The 
physics behind this is as follows: if outgoing particles are emitting soft 
(zero-momentum) gluons, they do not recoil. Thus, they must follow their 
classical straight-line trajectories. They can, however, change by
a phase. If this phase is to have the right gauge-transformation properties
to be part of a scattering amplitude, it must be a Wilson line operator. 
Given that eq.~(\ref{genfunc}) generates the soft function in the amplitude, 
another way to say the above is that the soft function is a vacuum expectation
value of Wilson lines, or:
\begin{equation}
{\cal S}\sim \left\langle 0\left|\prod_{k=1}^L\Phi_k\right|0\right\rangle,
\end{equation}
where $\Phi_k$ is a Wilson line operator evaluated along the
trajectory $x_k$ of particle $k$. Indeed, this is usually taken as the
definition of the soft function (see e.g.~\cite{Gardi:2009zv}). \\

Having introduced the generating functional for the soft gauge field theory,
we now proceed to use this in order to rederive abelian exponentiation. To this
aim, we introduce a crafty statistical physics approach, which is the subject
of the following section.

\subsection{The replica trick}
In the previous section, we introduced a generating functional for a QFT for 
the soft gauge field $A_\mu$, which generated soft photon subdiagrams. If we
consider the simple case of QED with no propagating fermions (so that soft 
photon graphs contain no fermion bubbles), the path integral over $A_\mu$ in
eq.~(\ref{genfunc}) is Gaussian, and can be carried out exactly. This would
lead immediately to abelian exponentiation, which we already encountered
in section~\ref{sec:abelian}. In the more complicated cases of QCD and QED with
fermion bubbles, it is no longer the case that the path integral in the 
generating functional can be performed exactly, and so we instead wish to 
introduce a method for deriving exponentiation of soft gauge boson 
diagrams which will be fully general. Specifically, we will use the {\it 
replica trick}, a technique borrowed from statistical physics 
(see e.g.~\cite{Replica}). We will see that this provides an elegant shortcut
to exponentiation, which bypasses the nontrivial combinatorics (e.g. the 
eikonal identity) that we encountered in section~\ref{sec:abelian}.\\

The argument proceeds as follows. Instead of the field theory defined by
eq.~(\ref{genfunc}), let us instead consider a theory with $N$ identical 
copies, or {\it replicas}, of the soft gauge field $A_\mu$. We can label these
replicas by $A_\mu^{(j)}$, where $1\leq j\leq N$. Importantly, we take these
replicas to be non-interacting. That is, a given replica does not interact 
with other replicas, so that the set of replica gauge fields is mutually 
independent. \\

\begin{figure}
\begin{center}
\scalebox{1.0}{\includegraphics{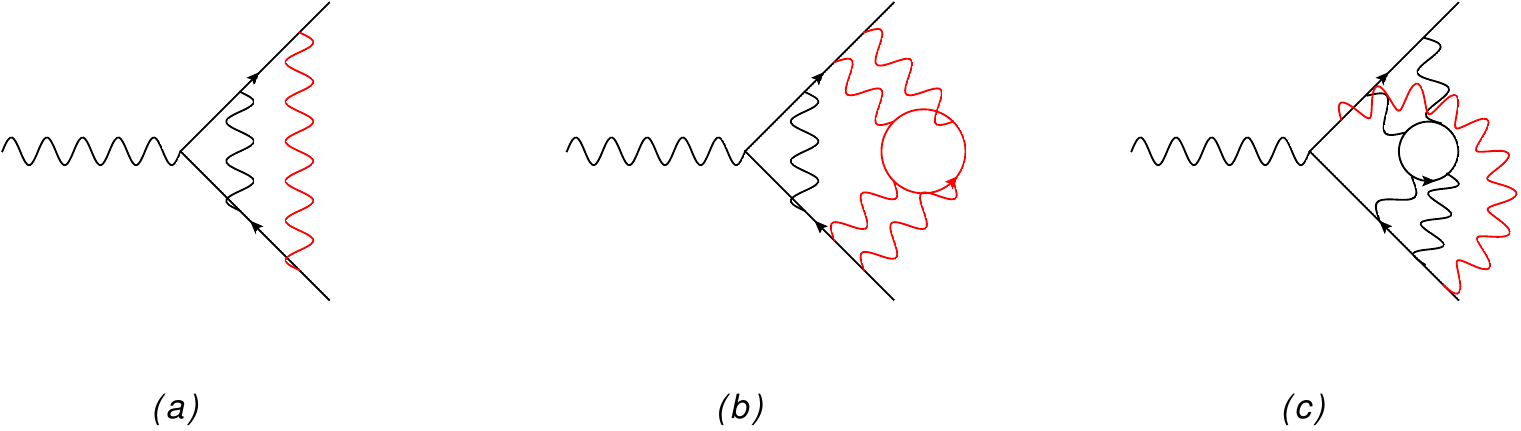}}
\caption{Examples of soft photon diagrams in the replicated theory of 
eq.~(\ref{genfuncrep}). Different colours represent potentially different 
replica numbers.}
\label{fig:repex}
\end{center}
\end{figure}

By generalising eq.~(\ref{genfunc}), the generating functional for the 
replicated theory is
\begin{equation}
Z_{\rm rep.}=\int{\cal D}A_\mu^{(1)}\,{\cal D}A_\mu^{(2)}\ldots{\cal D}
A_\mu^{(N)}\,\exp\left[i\sum_{j=1}^NS[A_\mu^{(j)}]\right]\,
\left[\prod_{k=1}^Le^{i\int dx_k\cdot A^{(1)}}\right]\ldots
\left[\prod_{k=1}^Le^{i\int dx_k\cdot A^{(N)}}\right].
\label{genfuncrep}
\end{equation}
Here there is a path integral over all possible field configurations
of each replica. The total action for the replicas is the sum of all
the individual replica actions, as follows from the fact that the
replicas are non-interacting. Furthermore, we now have a product of
Wilson line factors for each external line, where this product occurs
for each replica number from $1$ to $N$. We can simplify
eq.~(\ref{genfuncrep}) by noting that the Wilson line factors in an
abelian theory commute with each other, so that they can be combined
into a single exponential to give
\begin{equation}
Z_{\rm rep.}=\int{\cal D}A_\mu^{(1)}\,{\cal D}A_\mu^{(2)}\ldots{\cal D}
A_\mu^{(N)}\,\exp\left[i\sum_{j=1}^NS[A_\mu^{(j)}]\right]\,
\prod_{k=1}^L\exp\left[i\sum_{j=1}^N\int dx_k\cdot A^{(j)}\right].
\label{genfuncrep2}
\end{equation}
What do the diagrams generated by this theory look like? Ignoring the replica
index, the generating functional looks similar to the non-replicated theory,
and so diagrams in the replicated theory have similar topologies to the 
original soft photon theory. However, we have to take account of the fact that
each connected piece of a given diagram could have a different replica number. 
This is illustrated in figure~\ref{fig:repex}, which shows three examples 
of soft photon diagrams, dressing the hard interaction of 
figure~\ref{fig:amp}(a). 
The first example consists of two single soft photon emissions, each of which
may potentially correspond to a different replica (shown using different 
colours in the figure). We can also think about this in terms of assigning 
indices $i$ and $j$ to each photon, where $1\leq i,j\leq N$. In the second
example, there are two disconnected pieces in the soft photon subdiagram:
a single photon emission, and four photons connected to a fermion bubble. Each
of the separate connected pieces is constrained to have a single replica 
number. This follows from the fact that the action for each replica, 
$S[A_\mu^{(j)}]$, implicitly includes coupling to fermion bubbles etc. 
Thus, replicas cannot mix via matter bubbles, which would ruin the fact that 
the replicas are independent. The third example in the figure shows two 
similar connected pieces arranged in a different way, so that there is some 
sort of non-trivial kinematic entanglement between the different parts of the 
photon subgraph. This does not affect the fact, however, that the two 
connected pieces are completely separate from each other if the external lines
are removed (as defines the soft photon subgraph), and thus they may still 
have potentially different replica numbers.\\

The above discussion is perhaps rather unmotivated and mysterious! Let us now
see how we can use the replicated theory to prove abelian exponentiation. 
Firstly, note that the replicated generating functional of 
eq.~(\ref{genfuncrep}) is related to the generating functional of the original
theory, eq.~(\ref{genfunc}), by the relation
\begin{equation}
Z_{\rm rep.}=Z^N,
\label{ZN}
\end{equation}
where as usual $N$ is the total number of replicas. This follows 
from the fact that the replicas are non-interacting, so that one could choose
to rewrite eq.~(\ref{genfunc}) as a product of path integrals over each replica
$A_\mu^{(i)}$ individually, thus making the structure of eq.~(\ref{ZN}) 
manifest. We may now Taylor expand this equation in $N$ to give
\begin{equation}
Z_{\rm rep}=1+N\log Z+{\cal O}(N^2).
\label{Zexp}
\end{equation}
The left-hand side of this equation corresponds to a complete sum of Feynman
diagrams in the replicated theory. These will, in general, be $N$ dependent.
Equation~(\ref{Zexp}) then tells us that if we expand each Feynman diagram
as a function of $N$, the coefficient of the ${\cal O}(N)$ piece contributes
to the {\it logarithm} of $Z$ (the generating functional of the original 
theory). Note that this argument does not depend on $N$ being small - higher
order terms in $N$ contribute to higher powers of the logarithm of $Z$, 
which is irrelevant to our present logic. \\

It follows from the above discussion that if we find the ${\cal O}(N)$ part of
all diagrams in the replicated theory (let us call this $W$ for a given 
diagram), then the original generating functional is given by
\begin{equation}
Z=\exp\left[\sum_W W\right].
\label{Zexp2}
\end{equation}
Thus, $Z$ has a manifestly exponential form. This is not a particularly 
impressive statement in general - it is always true that $Z$ is the exponential
of its own logarithm! However, what the replica trick gives us is an explicit
procedure for determining the logarithm of $Z$; in other words, for determining
which diagrams exponentiate in the non-replicated theory. \\

To see how this works in this abelian case, consider for example the diagrams
in figure~\ref{fig:repex}(b) and~\ref{fig:repex}(c). These both have two 
separate connected pieces in the soft photon part of the diagram, and each of
these has a potentially different replica number. Given that there are $N$ 
choices for the first replica number and a further $N$ for the second, it
follows that diagrams with these topologies are ${\cal O}(N^2)$. Generalising
this argument, a diagram with $m$ separated connected pieces is 
${\cal O}(N^m)$. Such diagrams, by the above argument, do not contribute to 
the exponent of $Z$, as they have no part which is linear in $N$. The only
diagrams which are linear in $N$ have $m=1$ i.e. have only one connected piece.
We have thus derived, from the replica trick, that connected diagrams 
exponentiate in the soft photon theory~\footnote{This is in fact a special case
of a general result in quantum field theory (which can be found in many 
textbooks, albeit often in an appendix), that connected diagrams 
exponentiate.}! All disconnected diagrams, such as those shown in 
figure~\ref{fig:repex}(b) and (c), are obtained by the explicit exponentiation
of lower order diagrams. \\

Recalling that $Z$ generates subdiagrams in the full amplitude, namely those
parts of the graph that contribute to the soft function, we have derived the
result \\

\begin{center}
\fbox{${\cal S}=$exp[Connected soft photon subgraphs]}.\\
\end{center}

This is the abelian exponentiation result that we (partially) derived using
the eikonal identity in the previous section. Some further comments are as 
follows:
\begin{enumerate}
\item[(i)] The replica trick proof is much simpler than the eikonal identity
approach. Essentially, what the replica trick does is to hide the tricky 
combinatorics associated with building up the exponential series. Using a
traditional combinatoric approach would be even more complicated in non-Abelian
theories, although it can of course be done~\cite{Gatheral:1983cz,
Frenkel:1984pz,Sterman:1981jc,Gardi:2010rn,Mitov:2010rp}.
\item[(ii)] The replica trick works for any number of external lines. This will
be particuarly useful in the case of non-abelian gauge theories, in which 
results have traditionally been separated into the case of amplitudes with 
two coloured particles, and the general case of multiparton scattering.
\item[(iii)] The above result shows that {\it all} connected diagrams 
exponentiate, including those containing fermion bubbles. We would have had to
have proved this separately in the eikonal identity approach of 
section~\ref{sec:abelian} (although admittedly this is a minor 
modification~\cite{Yennie:1961ad}). 
\item[(iv)] In the replica trick analysis, we explicitly used the fact that
the Wilson line operators commute with each other, in order to combine their
exponents into a single exponential factor for each external line. A similar
result holds in quantum gravity, but not in non-abelian theories such as QCD.
The latter have colour matrices associated with the emission of a soft gluon,
which complicates the combinatorics of the path integral. In particular, the
standard textbook result of exponentiation of connected diagrams does not hold
for the field theory describing a soft gluon field (as we will see).  
\end{enumerate}
Having introduced the replica trick in this section, and described how this
can be used to derive exponentiation properties, we are now ready to explore
exponentiation in non-abelian theories. 

\section{Non-abelian exponentiation}
\label{sec:nonabel}
In the previous section, we introduced the replica trick and showed how this 
can be used to derive abelian exponentiation. The main idea was that the 
${\cal O}(N)$ part of Feynman diagrams in the replicated theory contributes to
the exponent of the soft function in the non-replicated theory. Thus, one has
an explicit procedure for calculating the exponent of the soft function 
directly. In non-abelian theories, the same overall procedure can be used,
but leads to a much richer structure in the exponent that is still not fully
understood. \\

Before examining non-abelian exponentiation in detail, it is worth making a
few remarks regarding the existing literature on this subject. These are given
in the following section, which can be skipped if necessary without affecting 
the flow of the argument from the previous section. However, what follows is 
useful for reasons of completeness, and for making contact with previous work 
on webs.

\subsection{Two parton vs. multiparton scattering}
\label{sec:2vsn}
\begin{figure}
\begin{center}
\scalebox{1.0}{\includegraphics{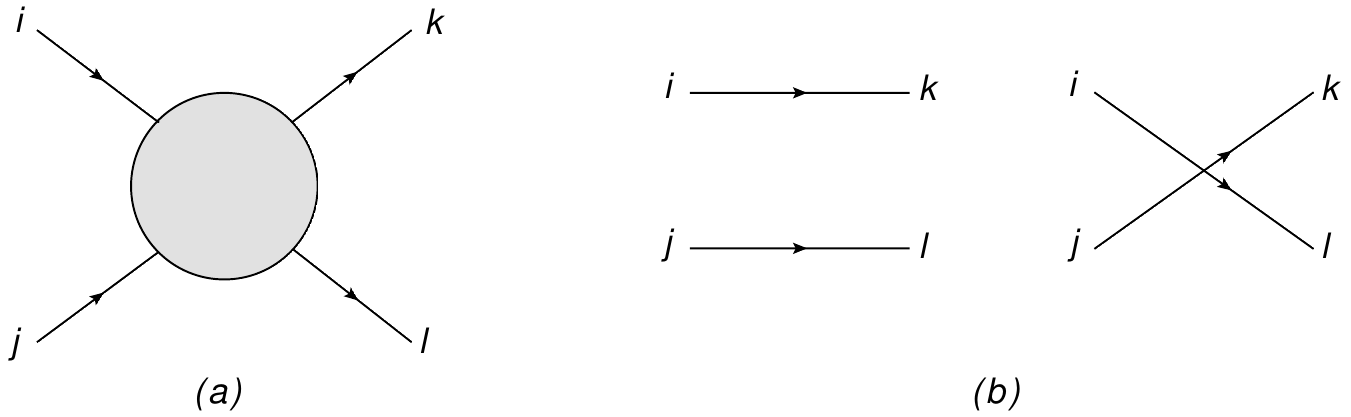}}
\caption{(a) Schematic depiction of $qq\rightarrow qq$ scattering, where the
labels denote colour indices; (b) The two independent ways of joining up the 
colour flow.}
\label{fig:qq}
\end{center}
\end{figure}

Historically, non-abelian exponentiation was first considered in the
case of scattering amplitudes involving only two coloured
particles~\cite{Gatheral:1983cz,Frenkel:1984pz,Sterman:1981jc} (plus
any number of colour singlet particles). Physical examples of these
processes include Drell-Yan production of a vector boson, Higgs boson
production via gluon fusion, deep inelastic scattering (in which an
electron collides with a proton, which subsequently breaks up), and
electron-positron annihilation into a quark-antiquark pair. In such
processes, each external quark or gluon has a colour index. The
special nature of amplitudes with only two coloured particles is that
there is only one possible colour flow at the hard interaction vertex:
the colour indices of the two quarks or gluons involved must join up
with each other (by colour conservation). In a general multiparton
scattering process, there are many colour indices joining up at the
hard interaction vertex, and consequently there are many different
ways of joining them up. Consider, for example, the case of
$qq\rightarrow qq$ scattering, as shown in figure~\ref{fig:qq}(a).
Whatever is happening in the hard interaction, it remains true that
there are only two ways of joining up the colour indices of the
incoming and outgoing quarks. These are shown in
figure~\ref{fig:qq}(b).
The scattering amplitude of this process will have four colour indices, as 
labelled in the figure~\ref{fig:qq}(a). The two ways of joining the colour
flow means that this amplitude must have the following form:
\begin{equation}
{\cal A}_{ijkl}={\cal A}_1\,\delta_{ik}\delta_{jl}+{\cal A}_2\,\delta_{il}
\delta_{jk},
\label{ampform}
\end{equation}
where $\delta_{ij}$ is the Kronecker delta symbol, and $\{{\cal A}_i\}$ are
kinematic coefficients which are independent of the colour index structure. 
We may rewrite this equation as
\begin{equation}
{\cal A}=\sum_I{\cal A}_I\,c_I,
\label{ampform2}
\end{equation}
where $I=1,2$ and the {\it colour tensors} $\{c_I\}$ are defined via
\begin{equation}
c_1=\delta_{ik}\delta_{jl},\quad c_2=\delta_{il}\delta_{jk}.
\label{cIdef}
\end{equation}
In other scattering amplitudes, there are more ways of joining up the
colour flow in general. However, amplitudes still have the generic
form of eq.~(\ref{ampform2}), with an appropriate set of colour
tensors. These are not uniquely defined - any basis obtained via a
linear transformation will do.  Having chosen a basis, we can work
with the components of the amplitude ${\cal A}_I$, and we see that
amplitudes are {\it vectors in colour-flow space}. The soft function
contains gluons which transfer colour from one external line to
another in general. Thus, it mixes the various colour flows present in
the hard interaction, and is matrix-valued in colour space. How such
matrices act on the colour structure of the hard interaction is an
interesting problem by itself - see
e.g. refs.~\cite{Sjodahl:2008fz,Sjodahl:2009wx}.\\

\begin{figure}
\begin{center}
\scalebox{1.0}{\includegraphics{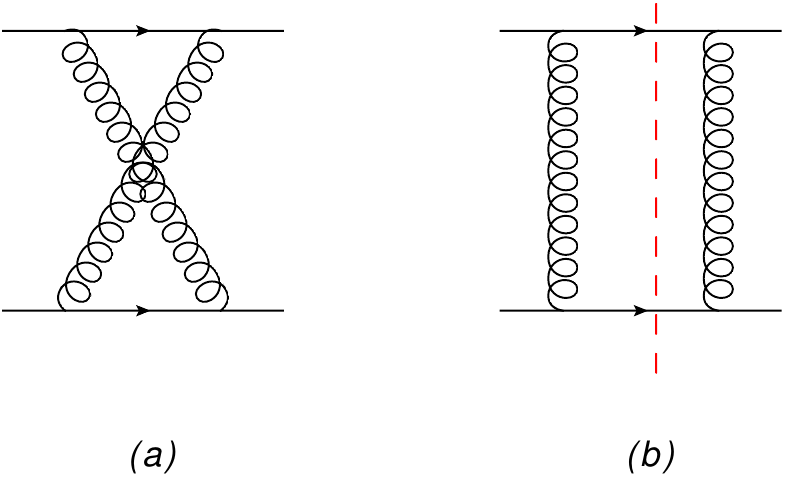}}
\caption{Examples of: (a) a two-eikonal line irreducible diagram; (b) a 
reducible diagram (the red line separates the two connected pieces).}
\label{fig:irred}
\end{center}
\end{figure}

Having introduced the above terminology, we can see why soft gluon
emission in the case of two hard external lines is simpler in general
than the multiline case. In the two line case, there is only one
colour tensor ($\delta_{ij}$, where $i$ and $j$ are the colour indices
of the two external lines). The soft function is then a $1\times1$
matrix, and different contributions to the soft function commute with
each other. This is in contrast to the general case of multiparton
scattering, in which the soft function is a non-commuting object in
colour flow space, and has implications for the structure of the
exponent of the soft function. It turns out that in the two-line case,
one can classify which diagrams appear to all orders in perturbation
theory using a simple topological criterion, namely that they are {\it
  two-eikonal line irreducible}. This means that one cannot disconnect
a given diagram by drawing a single line through both external lines,
and an example is given in figure~\ref{fig:irred}. Also shown is an
example of a reducible diagram.
In this case, figure~\ref{fig:irred}(b) would not enter the exponent of the 
soft function, whereas~\ref{fig:irred}(a) would. Recall that diagrams such as
the latter were not present in the exponent in the abelian case, so that even
in the two-line case QCD is already much more complicated than
QED~\footnote{This is only one complication of course. There are also soft 
gluon graphs in QCD which contain three and four gluon vertices, which do not
exist in QED.}. \\

\begin{center}
\parbox{5in}{{\bf Problem 3}. Draw all irreducible diagrams at 3 loop order.
}
\end{center}

The above results for the two line case were derived
in~\cite{Gatheral:1983cz,Frenkel:1984pz,Sterman:1981jc}, and indeed it
was a particular way of drawing two-eikonal line irreducible diagrams
in those papers that first led to the term {\it web}. We will not
present the derivation of this result here. It proceeds via a
generalisation of the eikonal identity of eq.~(\ref{eikid}), which
takes account of the fact that successive gluon emissions are
non-commuting (for a recent presentation,
see~\cite{Laenen:2010uz}). Rather, our aim is to present a formalism
for non-abelian exponentiation, which is set up for the general case
of multiparton scattering. The special properties of two-line
scattering will then emerge as a special case. The following sections
are based heavily on~\cite{Gardi:2010rn}, but note that multiline webs
have also been considered from a different point of view
in~\cite{Mitov:2010rp}. For a discussion of how these two approaches
are related, see appendix A of~\cite{Gardi:2011wa}. More recently,
another approach has also
appeared~\cite{Vladimirov:2014wga,Vladimirov:2015fea}.

\subsection{The replica trick in a non-abelian theory}
\label{sec:nonabreplica}
We now return to elucidating the structure of non-abelian exponentiation.
In the abelian case of section~\ref{sec:path}, we used the replica trick
to elegantly derive the exponentiation structure of the soft function, namely
that connected soft photon subdiagrams enter the exponent. This strongly
suggests that we should use the replica trick also in the non-abelian case. 
Let us then try this and see what happens.\\

First we need to write down the generating functional for the soft gluon 
theory, whose diagrams are the soft gluon subdiagrams of a non-abelian
scattering amplitude. This is a straightforward generalisation of 
eq.~(\ref{genfunc}), and the result is
\begin{equation}
Z=\int{\cal D}A_\mu e^{iS[A_\mu]}\prod_{k=1}^L{\cal P}
e^{i\int dx_k^\mu A_\mu^k}.
\label{genfuncnab}
\end{equation}
This has a path integral over the soft gluon field, with a weighting
factor containing the classical action. For ease of notation we use
the same symbol $S[A_\mu]$ as in the QED case. However, the non-abelian
action will contain three and four gluon vertices etc. in addition to
the propagator and couplings to matter bubbles. Each external line, as
before, has a Wilson line operator associated with it. This is similar
to the QED Wilson line, except for the fact that the gauge field
$A_\mu^k=A_\mu^a{\bf T}_k^a$ is matrix-valued, where ${\bf T}_k$ is a
colour matrix acting on the partonic colour indices of the $k^{\rm
  th}$ line.  We do not write these colour indices explicitly in
eq.~(\ref{genfuncnab}), but we do add an index $k$ to the soft gauge
field to remind us that this lives in the colour index space of the
$k^{\rm th}$ parton line. Given that the exponent of the Wilson line
operator is matrix-valued, the exponential itself is ambiguous. Upon
carrying out a Taylor expansion, it is not clear how to order the
colour matrices in the correct fashion. This is corrected by the
operator ${\cal P}$ in eq.~(\ref{genfuncnab}), which stands for {\it
  path ordering}. This operator is defined such that, in a given term
in the expansion of the exponential, the colour matrices are ordered
along the line according to the sequential ordering of gluon
emissions. Parametrising the contour of a given line according to
eq.~(\ref{xkdef}), for example, this means that the colour matrices
will be ordered according to the order of the $t$ values at which
gluons are emitted. This is a somewhat subtle point that is absent in
the case of an abelian gauge theory, where no non-commuting charges
appear. \\

Applying the replica trick, one may consider the theory
\begin{align}
Z_{\rm rep.}&=\int{\cal D}A_\mu^{(1)}\ldots{\cal D}A_\mu^{(N)}
\exp\left[i\sum_{j=1}^NS[A_\mu^{(j)}]\right]\notag\\
&\quad\times\left[\prod_{k=1}^L{\cal P}\exp
\left(i\int dx_k^\mu\,A^{k(1)}_\mu\right)\right]\ldots\left[\prod_{k=1}^L
{\cal P}\exp\left(i\int dx_k^\mu\,A^{k(N)}_\mu\right)\right],
\label{genfuncrepnab}
\end{align}
where we have again taken $N$ independent (mutually non-interacting) copies of
the gauge field. At this point in the abelian case, we collected together all
Wilson line exponents to make a single exponential. We cannot do that in this 
case, due to the fact that the exponents of the Wilson line factors do not 
commute, if these act on the same external line. We can, however, commute 
Wilson line operators which act on {\it different} external lines. These have
completely different colour indices, so that matrices on different lines
commute simply by not living in the same colour algebra. This allows us to 
gather together the contributions from each external line separately in 
eq.~(\ref{genfuncrepnab}), to get
\begin{align}
Z_{\rm rep.}&=\int{\cal D}A_\mu^{(1)}\ldots{\cal D}A_\mu^{(N)}
\exp\left[i\sum_{j=1}^NS[A_\mu^{(j)}]\right]\prod_{k=1}^L\left[
\prod_{j=1}^N{\cal P}\exp
\left(i\int dx_k^\mu\,A^{k(j)}_\mu\right)\right].
\label{genfuncrepnab2}
\end{align}
To recap, the first product in this formula is over all external lines. Then,
each external line factor has a product of Wilson line operators acting on it,
one for each replica. Furthermore, the replica number in this product increases
from left to right, from 1 to $N$. \\

We cannot immediately read off the Feynman diagrams for the soft gauge
field from eq.~(\ref{genfuncrepnab2}), as this contains a product of
path-ordered exponentials for each external line. Were we inclined to
rewrite this as a single path-ordered exponential, we would then be
able to analyse the diagrams in more detail, by analogy with the case
of the original theory. In general combining a number of exponentials
involving non-commuting objects involves multiple applications of the
Baker-Campbell-Hausdorff formula (see
e.g. refs.~\cite{Gardi:2013ita,Vladimirov:2014wga,Vladimirov:2015fea}
for approaches to exponentiation that build on this idea). However, we
can circumvent this in a crafty way by exploiting the fact, mentioned
above, that the replica number increases sequentially from $1$ to $N$
in the product of Wilson line operators associated with each external
line. Thus, one may write this product as
\begin{equation}
\prod_{j=1}^N{\cal P}\exp\left(i\int dx_k^\mu\,A_\mu^{k(i)}\right)
={\cal R}{\cal P}\exp\left[i\sum_{i=1}^N\int dx_k^\mu\,A_\mu^{k(i)}\right].
\label{wilprod}
\end{equation}
Here ${\cal R}$ is a {\it replica-ordering operator}. Given a product of
colour matrices on a given line, it ensures that the replica number 
increases along the line. We can define its action more formally by 
considering two colour matrices acting on line $k$. Denoting by 
${\bf T}^{(i)}_k$ a colour matrix acting on line $k$ and associated with a 
gluon with replica number $i$, one has
\begin{equation}
{\cal R}\left[{\bf T}_k^{(i)}\,{\bf T}_k^{(j)}\right]=\left\{\begin{array}{c}
{\bf T}_k^{(i)}\,{\bf T}_k^{(j)}\quad i\leq j\\
{\bf T}_k^{(j)}\,{\bf T}_k^{(i)}\quad i>j\end{array}\right..
\label{repdef}
\end{equation}
That is, ${\cal R}$ does nothing if the replica numbers associated
with colour matrices are already increasing (or the same) along a
given line, but reorders them if this is not the case. It is
straightforward to generalise this definition to higher numbers of
matrices. The ${\cal R}$ operator then acts on each term in the
expansion of the exponential on the right-hand side of
eq.~(\ref{wilprod}), and reorders the colour matrices such that the
replica numbers in each term are in increasing order. Then the
right-hand side agrees with the expansion of the multiple exponentials
on the left-hand side which, if carried out, will have the replica
numbers similarly ordered in every term. Note that the notation of
eq.~(\ref{wilprod}) implies that one first path orders (according to
${\cal P}$), and then replica orders (according to ${\cal R}$). In
cases where these orderings are different ${\cal R}$ overrides ${\cal
  P}$. \\

Substituting eq.~(\ref{wilprod}) into eq.~(\ref{genfuncrepnab2}), we may 
rewrite the latter as 
\begin{align}
Z_{\rm rep.}&=\int{\cal D}A_\mu^{(1)}\ldots{\cal D}A_\mu^{(N)}
\exp\left[i\sum_{j=1}^NS[A_\mu^{(j)}]\right]{\cal R}\prod_{k=1}^L\left[
{\cal P}\exp\left(i\sum_{j=1}^N\int dx_k^\mu\,A^{k(j)}_\mu\right)\right].
\label{genfuncrepnab3}
\end{align}
Here, to simplify notation, we have defined an overall replica-ordering
operator ${\cal R}$ which is understood to act appropriately on each 
line $k$. Note that this now looks more similar to the QED case of
eq.~(\ref{genfuncrep2}), apart from the colour-specific details (which are also
present in the non-replicated QCD theory), and also the presence of the 
${\cal R}$ operator.\\

Consider now a particular soft gluon diagram $D$ with a given topology. In the
original (non-replicated) theory, this has the general form
\begin{displaymath}
{\cal F}(D)C(D),
\end{displaymath}
where $C(D)$ is the colour factor of the graph (collecting all the soft gluon
colour matrices on each external line), and ${\cal F}(D)$ contains all the 
kinematic dependence. In the replicated theory, the same diagram topology
will have the similar form
\begin{displaymath}
{\cal F}(D)\hat{C}(D),
\end{displaymath}
where ${\cal F}(D)$ is the same as in the original theory (as in the QED case,
the kinematics of a given diagram does not care about the assignment of the 
replica numbers). However, the colour factor $\hat{C}(D)$ is {\it not the same}
as in the original theory. Firstly, it depends on the number of replicas $N$.
Secondly, colour factors are different in the replicated theory due to the 
fact that colour matrices, which obey path ordering only in the original 
theory, get reordered by the ${\cal R}$ operator. The details of this 
reordering depend on the particular assignment of replica numbers in a given
graph. To find the total colour factor in the replicated theory, one must also
sum over the total number of replica assignments. The whole procedure is 
perhaps best illustrated by example. \\

\begin{figure}
\begin{center}
\scalebox{0.8}{\includegraphics{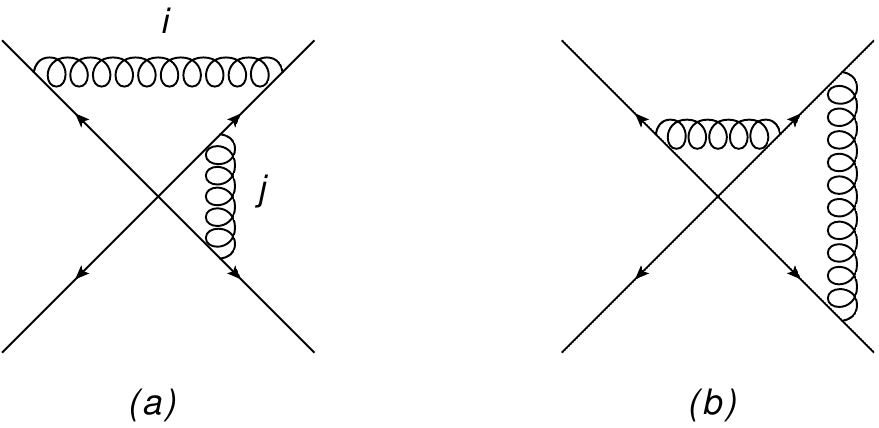}}
\caption{Example two-loop diagrams, where $i$ and $j$ denote replica numbers.}
\label{fig:2loop}
\end{center}
\end{figure}

Consider the two diagrams at two-loop order shown in figure~\ref{fig:2loop}.
Here we have considered an amplitude with four external lines, and shrunk the
hard interaction to a point. 
In the replicated theory, the two diagrams in the figure give contributions
to the (replicated) amplitude
\begin{equation}
{\cal A}_a={\cal F}(a)\hat{C}(a),\quad {\cal A}_b={\cal F}(b)\hat{C}(b)
\label{repamp}
\end{equation}
respectively. To find the total colour factor of diagram (a), one must
consider all possible assignments of replica number $i$ and $j$, as
these will potentially contribute different amounts to the total,
depending on the action of the replica ordering operator ${\cal
  R}$. Given that the action of this operator only depends upon
whether replica indices are greater or less than each other (or the
same), we need only consider three cases: $i=j$, $i<j$ and
$i>j$. These are given in the first column of table~\ref{tab:rep2},
and the third column gives the total number of replica number
assignments corresponding to each hierarchy of $i$ and $j$. In the
second column, we note the $N$-independent part of the contribution to
the colour factor for each diagram. Considering first the case of
$i=j$, this is simply the colour factor $C(a)$ of (a) in the original
theory, as no replica reordering takes place if the replica indices
are the same. If $i<j$, the gluons are not ordered in terms of
increasing replica number on the upper right-hand line (where,
according to the arrows to designate our choice of momentum flow, we
are taking the direction of the Wilson line as increasing away from
the hard interaction vertex). Then the colour matrices get reordered,
which produces the conventional colour factor $C(b)$ of diagram
(b). This is the reason why we have had to consider two diagrams at
once - they mix with each other due to the ${\cal R}$
operator. Finally, we must consider the case $i>j$. In this case the
replica ordering constraint is already satisfied, and thus the
$N$-independent part of the contribution to the colour factor is
$C(a)$.\\
\begin{table}
\begin{center}
\begin{tabular}{c|c|c}
$i$, $j$ & Replica-ordered colour factor & Multiplicity\\
\hline
$i=j$ & $C(a)$ & $N$\\
$i<j$ & $C(b)$ & $N(N-1)/2$\\
$i>j$ & $C(a)$ & $N(N-1)/2$
\end{tabular}
\caption{Replica trick analysis of the diagrams of figure~\ref{fig:2loop}, as
described in the text.}
\label{tab:rep2}
\end{center}
\end{table}

Putting things together, the total colour factor in the replicated theory is
\begin{align}
\hat{C}(a)&=NC(a)+\frac{N(N-1)}{2}\left[C(a)+C(b)\right]\notag\\
&=N\left[\frac{C(a)-C(b)}{2}\right]+{\cal O}(N^2).
\label{colfaca}
\end{align}
Here we have formed the total by multiplying together the results of the second
and third columns in table~\ref{tab:rep2}, and summing these up. In the second
line we have taken the ${\cal O}(N)$ part of the colour factor, anticipating 
that this will be relevant in what follows.\\

One may carry out a similar analysis for diagram (b) in figure~\ref{fig:2loop},
and the total colour factor in the replicated theory is found to be
\begin{equation}
\hat{C}(b)=N\left[\frac{C(b)-C(a)}{2}\right]+{\cal O}(N^2).
\label{colfacb}
\end{equation}
We may now use the replica trick argument from section~\ref{sec:path}, which 
said that the contribution of a given diagram $D$ to the exponent of the 
generating functional in the non-replicated theory is given by the 
${\cal O}(N)$ part of this diagram in the replicated theory. Thus, we see that
both diagrams (a) and (b) contribute to the exponent of the soft function in
the original theory, and they do so in the combination (taking the coefficients
of the ${\cal O}(N)$ terms in eq.~(\ref{colfaca}, \ref{colfacb}))
\begin{equation}
\left(\frac{C(a)-C(b)}{2}\right){\cal F}(a)
+\left(\frac{C(b)-C(a)}{2}\right){\cal F}(b).
\label{expab}
\end{equation}
Note that this can be rewritten in the matrix form
\begin{equation}
\left(\begin{array}{c}{\cal F}(a)\\{\cal F}(b)\end{array}\right)^T
\frac{1}{2}\left(\begin{array}{rr}1&-1\\-1&1\end{array}\right)
\left(\begin{array}{c}C(a)\\C(b)\end{array}\right).
\label{expab2}
\end{equation}
This is an interesting structure, as the vectors on the left and right-hand
side (defined in the space of diagrams) involve properties of the original
theory, as they should now that we have returned from the replicated theory.
The matrix in the middle consists of constant numbers, and describes a mixing
between the colour and kinematic degrees of freedom of soft gluon diagrams in
the exponent of the soft function. This structure is in fact entirely general,
as we describe in the following section.

\begin{figure}
\begin{center}
\scalebox{0.8}{\includegraphics{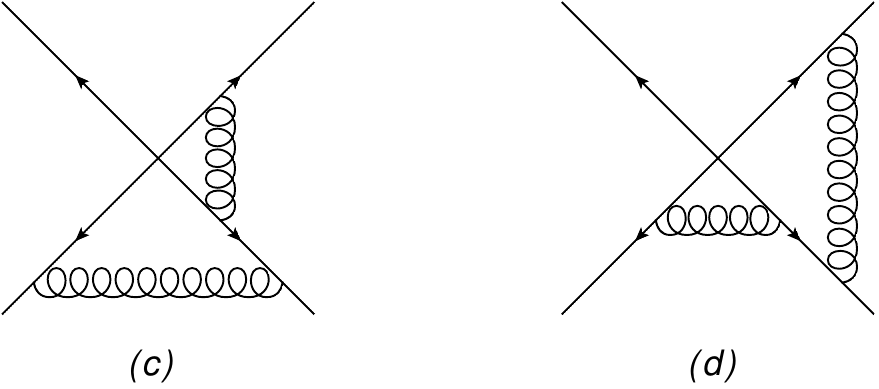}}
\caption{Diagrams considered in problem 4.}
\label{fig:cd}
\end{center}
\end{figure}

\begin{center}
\parbox{5in}{{\bf Problem 4}. Without detailed calculation, write down the
contribution to the soft gluon exponent of the diagrams in figure~\ref{fig:cd}.
}
\end{center}

\subsection{General structure of multiparton webs}
\label{sec:genstruc}
In the previous section, we examined a particular set of diagrams that mix
under the action of the ${\cal R}$ operator. Both diagrams were found to 
contribute to the exponent of the soft function, but with a mixing matrix
that entangles the colour and kinematic degrees of freedom. To see how this
works in general, note first that the two diagrams of figure~\ref{fig:2loop} 
are related by gluon permutations on the external lines. In this simple case, 
there is only one non-trivial permutation possible, on the upper right-hand 
line. Furthermore, this is a closed set: acting with more permutations will
not generate any additional diagrams. In general, given a diagram $D$
involving arbitrary numbers of gluon attachments on arbitrary external lines
(where these gluons may also be connected away from the external lines by 
fermion bubbles or multiple gluon vertices), we can always form such a closed
set of diagrams related by gluon permutation operations. A further example is
given in figure~\ref{fig:17-20}. It is clear that a given diagram can only
mix with diagrams from the closed set to which it belongs. This is because the
source of the mixing is the ${\cal R}$ operator in the replicated theory, which
is itself a form of gluon permutation operator, transforming between diagrams
in such a closed set. \\

Now let $W$ denote such a closed set. A given diagram $D\in W$ has a total 
colour factor $\hat{C}$ in the replicated theory. This gives a contribution 
to the soft function in the original theory a quantity
\begin{displaymath}
\sum_D{\cal F}(D)\tilde{C}(D),
\label{expcont}
\end{displaymath}
where we define $\tilde{C}(D)$ to be the coefficient of the ${\cal O}(N)$
part of $\hat{C}(D)$. This is the part of the colour factor in the replicated
theory that contributes to the exponent of the soft function in the original
theory. For this reason, $\tilde{C}(D)$ is referred to as the
{\it exponentiated colour factor} (ECF) of diagram $D$~\cite{Gardi:2010rn}.
In general, $\tilde{C}(D)$ will be a linear superposition
of the original colour factors $C(D)$ (this superposition can in principle 
be found by applying the replica trick algorithm as outlined in the two loop
example). Thus, one may write
\begin{equation}
\tilde{C}(D)=\sum_{D'}R_{DD'}C(D')
\label{ctildec}
\end{equation}
for some constants $R_{DD'}$. Let us now call the set of closed diagrams $W$ 
a {\it web}, and $R_{DD'}$ elements of a {\it web mixing matrix}. This can be
summarised as follows:\\

\begin{center}
\parbox{5in}{In non-abelian theories, webs are closed sets of diagrams 
related by gluon permutations on the external lines. Each web is 
described by a web mixing matrix, which encodes how colour and 
kinematic degrees of freedom mix in the exponent of the soft 
function.}
\end{center}

It is perhaps clear that we want to consider a closed set of diagrams
in some sense as a single web, given that these diagrams are mutually
entangled.  In fact, the motivation to call such sets webs goes
further than this, as we will see in section~\ref{sec:renorm}. The
general structure of the soft function in multiparton scattering can
then be written schematically as
\begin{equation}
{\cal S}=\exp\left[\sum_W\sum_{D,D'}{\cal F}(D)R^{(W)}_{DD'}C(D')\right],
\label{Sstruc}
\end{equation}
where the first sum is over all possible webs $W$ with mixing matrices
$R^{(W)}_{DD'}$, and the additional sums are over the diagrams in each web.\\

It is clear, then, that the general study of webs in multiparton scattering is
dominated by the study of web mixing matrices. These encode a potentially 
huge amount of physics! Namely, web mixing matrices encode how colour and 
kinematic information is entangled to all orders in perturbation theory, and 
the hope is that significant insights into the physics of non-abelian gauge
theories can be obtained by studying web mixing matrices. 

\subsection{A three loop example}
\label{sec:3loop}
\begin{figure}
\begin{center}
\scalebox{1.0}{\includegraphics{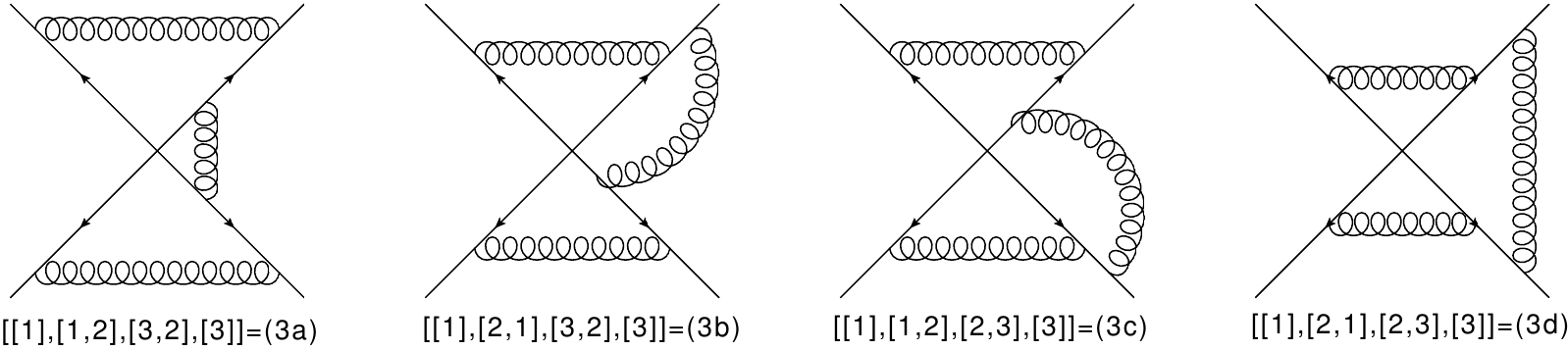}}
\caption{An example of a closed set of diagrams at three loop order.}
\label{fig:17-20}
\end{center}
\end{figure}

In order to clarify the above statements, it is useful to consider a three
loop example of applying the replica trick algorithm, to complement the two
loop example from earlier. Consider the set of diagrams shown in 
figure~\ref{fig:17-20}, and let us calculate the exponentiated colour factor 
$\tilde{C}$ of the diagram labelled (3a) in the figure. 
As in the two loop case, we must consider all possible assignments of replica
numbers. Labelling the upper, right-hand and lower gluons with replica numbers
$i$, $j$ and $k$ respectively, we must thus consider all possible hierarchies
of $i$, $j$ and $k$. These are given in table~\ref{tab:3loop}, which may be
compared to the two loop example of table~\ref{tab:rep2}.
\begin{table}
\begin{center}
\begin{tabular}{c|c|c|c}
$i$, $j$, $k$ & Replica-ordered colour factor & Multiplicity \\
\hline
$i=j=k$&$C(3a)$ & $N$\\
$i=j,k>i$&$C(3a)$ & $N(N-1)/2$ \\
$i=j,k<i$&$C(3c)$ & $N(N-1)/2$ \\
$i=k,j>i$&$C(3d)$& $N(N-1)/2$ \\
$i=k,j<i$&$C(3a)$& $N(N-1)/2$ \\
$j=k,i>j$&$C(3a)$& $N(N-1)/2$ \\
$j=k,i<j$&$C(3b)$& $N(N-1)/2$ \\
$i<j<k$&$C(3b)$& $N(N-1)(N-2)/6$ \\
$i<k<j$&$C(3d)$& $N(N-1)(N-2)/6$ \\
$j<i<k$&$C(3a)$& $N(N-1)(N-2)/6$ \\
$j<k<i$&$C(3a)$& $N(N-1)(N-2)/6$ \\
$k<i<j$&$C(3d)$& $N(N-1)(N-2)/6$ \\
$k<j<i$&$C(3c)$& $N(N-1)(N-2)/6$ 
\end{tabular}
\caption{Replica trick analysis for diagram (3a), as shown in 
figure~\ref{fig:17-20}. The replica indices $i$, $j$ and $k$ label the gluons 
on the upper, right-hand and lower parts of the diagram respectively.}
\label{tab:3loop}
\end{center}
\end{table}
As in the previous example, for each hierarchy of replica numbers, the 
$N$-independent part of the contribution to the total colour factor 
$\hat{C}(3a)$ is found by applying the ${\cal R}$ operation to the diagram, and
matching the result to one of the other diagrams in the web. One must then 
multiply each result by the number of distinct replica number assignments
that are consistent with the given hierarchy. These multiplicities are given in
the third column of table~\ref{tab:3loop}. The total colour factor in the 
replicated theory is then found to be
\begin{align}
\hat{C}(3a)&=NC(3a)+\frac{N(N-1)}{2}\left[3C(3a)+C(3b)+C(3c)+C(3d)\right]
\notag\\
&\quad+\frac{N(N-1)(N-2)}{6}\left[2C(3a)+2C(3d)+C(3b)+C(3c)\right].
\label{Chat3a}
\end{align}
The exponentiated colour factor in the original theory is the coefficient of
the ${\cal O}(N)$ part of this, which is
\begin{align}
\tilde{C}(3a)&=C(3a)\left[1-\frac{3}{2}+\frac{2}{6}\right]+C(3b)\left[
-\frac{1}{2}+\frac{1}{6}\right]+C(3c)\left[-\frac{1}{2}+\frac{1}{6}\right]
+C(3d)\left[-\frac{1}{2}+\frac{2}{6}\right]\notag\\
&=\frac{1}{6}\left[C(3a)-C(3b)-C(3c)+C(3d)\right].
\label{Chat3a2}
\end{align}
From the definition of the web mixing matrix of eq.~(\ref{ctildec}), this means
that the first row of the web mixing matrix is $\frac{1}{6}(1,-1,-1,1)$.
In fact, the full matrix turns out to be~\cite{Gardi:2010rn}
\begin{equation}
R_{DD'}=\frac{1}{6}\left(\begin{array}{rrrr}1&-1&-1&1\\
-2&2&2&-2\\-2&2&2&-2\\1&-1&-1&1\end{array}\right).
\label{R17-20}
\end{equation}

\begin{center}
\parbox{5in}{{\bf Problem 5}. Verify this result by calculating the remaining
rows of the matrix.
}
\end{center}

As the above problem demonstrates, calculation of web mixing matrices
in any particular case is straightforward, but tedious. The algorithm
can be - and indeed has been - automated. Many more examples of web
mixing matrices can be found in~\cite{Gardi:2010rn,Gardi:2011yz}, and
were obtained by computer.

\subsection{General properties of web mixing matrices}
\label{sec:props}
We have now seen something of the general structure of multiparton webs, and
in particular the fact that colour and kinematic parts of soft gluon diagrams
mix according to web mixing matrices. We have seen examples of how to calculate
these matrices in particular examples. Note in particular that this is a
purely combinatoric exercise. This suggests, 
broadly speaking, two different ways to investigate the further properties of
webs:
\begin{enumerate}
\item One may surmise the general structure of web mixing matrices by studying
the {\it physics} of soft gluon graphs (e.g. the known properties of colour
factors when the number of colours becomes large, or the need for infrared
singularities to be consistent with renormalisation of the vertex at which all
the external lines meet).
\item One may study general properties of web mixing matrices using {\it pure
mathematics}, namely the combinatorics underlying the replica trick algorithm.
The hope is then that established properties of the web mixing matrices can be
translated into (previously unknown) physics.
\end{enumerate}
Both of these are currently under investigation, and here we give a
taste of some general features of web mixing matrices that have so far
been established. \\

Firstly, we know that web mixing matrices have {\it zero sum rows}. That is,
any mixing matrix satisfies
\begin{equation}
\sum_{D'}R_{DD'}=0,
\label{zerosum}
\end{equation}
where the sum is over any row (labelled by $D$). One may verify this
for the two examples given above in eqs.~(\ref{expab2})
and~(\ref{R17-20}). This result means that one may redefine the ECF of
all diagrams in a web by a common constant. Another way of looking at
this is that the contribution to the colour factor of a given diagram
which is completely symmetric under interchanges of diagrams in the
web does not enter the exponent of the soft function. The zero sum row
rule has also been extended in~\cite{Gardi:2011wa}, where it is shown
that certain subrows also sum to zero, corresponding to subsets of the
web corresponding to planar and non-planar diagrams. Furthermore, a
similar sum rule involving columns has been observed, involving extra
weight factors~\cite{Gardi:2011yz}.\\

Secondly, all web mixing matrices are {\it idempotent}, i.e.
\begin{equation}
\sum_{E}R_{DE}R_{ED'}=R_{DD'}.
\label{idemp}
\end{equation}
This means that web mixing matrices are {\it projection operators},
having eigenvalues 0 and 1 (each with some multiplicity). These
properties are crucial to the cancellation of infrared singularities
in the exponent of the soft function, as has been explored
in~\cite{Gardi:2011yz,Gardi:2013saa,Falcioni:2014pka}. \\

Both of the above properties have been proven to be completely general,
including webs whose diagrams contain three and four gluon 
vertices~\cite{Gardi:2011wa}. The proof of the idempotence property involves
applying the replica trick twice in succession, and arguing that this is 
equivalent to applying it once (i.e. replicating a theory $M$ then $N$
times is equivalent to replicating it $MN$ times). The proof of the zero sum
row property uses results from enumerative combinatorics, in particular the 
counting of partitions of sets via Stirling numbers of the second kind.\\

\begin{figure}
\begin{center}
\scalebox{0.7}{\includegraphics{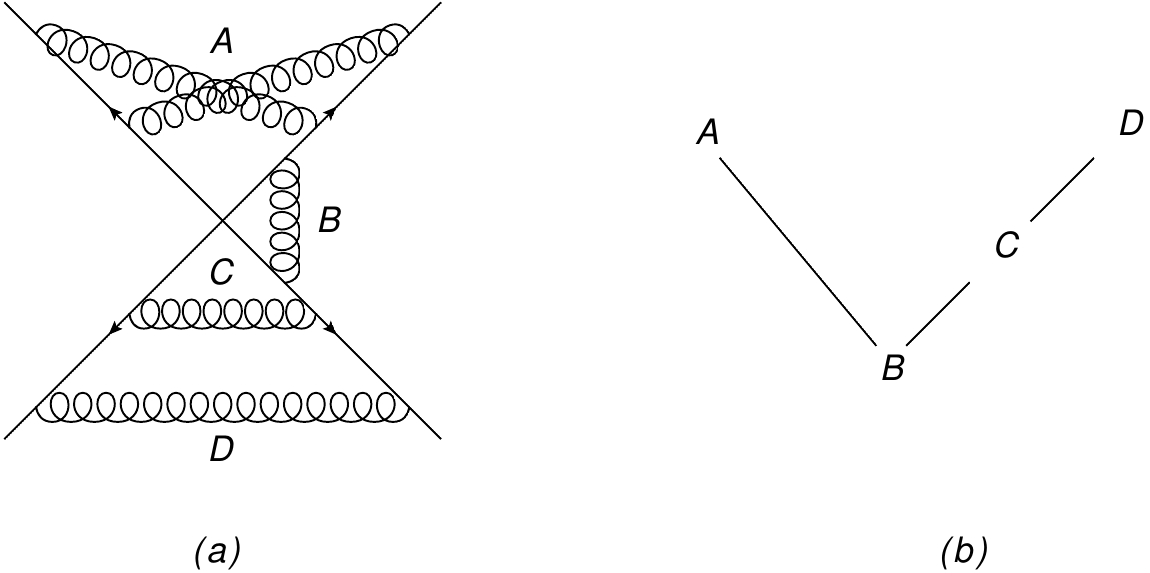}}
\caption{(a) Example web diagram; (b) Hasse diagram corresponding to the
poset generated by the diagram in (a).}
\label{posetex2}
\end{center}
\end{figure}

Recently, both the mathematics and physics behind web mixing matrices
have been made clearer. In ref.~\cite{Dukes:2013wa}, the combinatorics
of mixing matrices was related to that of partially ordered sets, or
{\it posets}. As the name suggests, a poset is a set of objects $X$
endowed with an ordering operation $\leq$ possessing the following
properties:
\begin{enumerate}
\item {\it Reflexivity}. If $x\in X$ then $x\leq x$.
\item {\it Antisymmetry}. If $x\leq y$ and $y\leq x$ then $x=y$.
\item {\it Transitivity}. If $x\leq y$ and $y\leq z$, $x\leq z$.
\end{enumerate}
This ordering operation applies to pairs of elements in the set $X$,
but such that not all pairs are necessarily ordered with respect to
each other (hence the ``partially'' in ``partially ordered set'').
The relation to web diagrams is as follows: every web diagram $D$ can
be identified with a poset. The elements of the poset are the various
irreducible subdiagrams that occur in the diagram $D$. The ordering
operation $\leq$ is defined as follows: given two irreducible
subdiagrams $x$ and $y$, $x\leq y$ if $x$ lies closer to the origin
than $y$, in the sense that one cannot pull $y$ into the origin
without also pulling in $x$. As an example, consider the diagram shown
in figure~\ref{posetex2}(a). In this diagram, the two gluons labelled
$A$ form an irreducible subdiagram: the fact that they are crossed
means that one cannot shrink either gluon to the origin independently
of the other one. The single gluon exchanges labelled $B$, $C$ and $D$
are also irreducible subdiagrams. Then the structure of the diagram is
encoded by the information that $B$ must be shrunk before $A$, $C$ or
$D$, and $C$ must be shrunk before $D$. This corresponds to the {\it
  Hasse diagram} shown in figure~\ref{posetex2}(b): a diagram whose
vertices represent the elements of the poset, and whose edges denote
an ordering between two elements. \\

\begin{center}
\parbox{5in}{{\bf Problem 6}. Draw the Hasse diagrams for the web of
  figure~\ref{fig:17-20}, and hence show that these correspond to a
  (possibly kinked) chain of linked vertices.}
\end{center}

The language of posets and Hasse diagrams is a powerful
one. Reference~\cite{Dukes:2013wa} uses it to prove a number of
general results, including results for the the rank of the mixing
matrix for three given families of web, for any number of
gluons. Reference~\cite{Dukes:2013gea} carries this even further. For
two particular web families, solutions are given for the web mixing
matrix for any number of gluons. This hints at the possibility that
the combinatorics of web mixing matrices could be completely
classified, and that it would be possible to go straight from web
diagrams to the mixing matrix, without having to apply the replica
trick algorithm.\\

\begin{figure}
\begin{center}
\scalebox{0.6}{\includegraphics{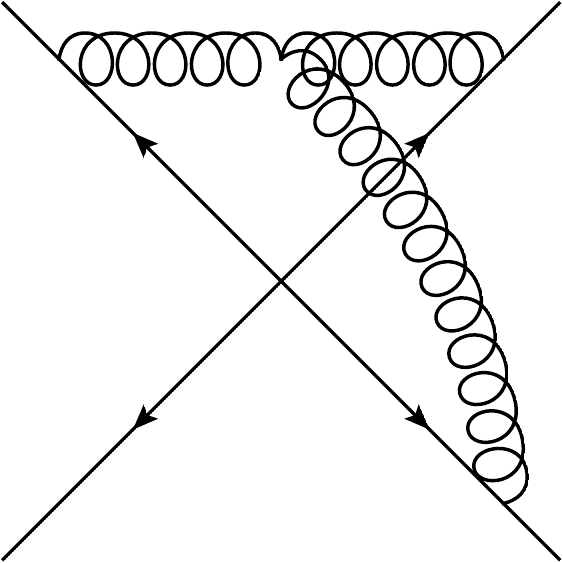}}
\caption{Connected gluon graph connecting three lines at two loops.}
\label{fig:connected}
\end{center}
\end{figure}

The physical content of web mixing matrices has been clarified in
ref~\cite{Gardi:2013ita}. As explained above, mixing matrices act as
projection operators, and thus have eigenvalues
$\lambda\in\{0,1\}$. Associated with each unit eigenvalue is a
combination of kinematic factors of diagrams, accompanied by a
superposition of colour factors. These colour factor combinations have
the general property that they look like colour factors of {\it
  connected} gluon subgraphs. To illustrate this, we may use the
two-loop web of figure~\ref{fig:2loop}. From eq.~(\ref{expab}), we see
that the contribution to the exponent from this web is
\begin{equation}
\frac{1}{2}\left(C(a)-C(b)\right)\left({\cal F}(a)-{\cal F}(b)\right),
\label{expabconn}
\end{equation}
and the fact that there is only one combination of kinematic factors
(with accompanying colour factor) is because the web mixing matrix in
eq.~(\ref{expab2}) is of rank 1, i.e. has only one unit
eigenvalue. The colour factor appearing in eq.~(\ref{expabconn}) is
\begin{equation}
{\bf T}_1^a\,{\bf T}_2^b\,{\bf T}_2^a\,{\bf T}_3^b
-{\bf T}_1^a\,{\bf T}_2^a\,{\bf T}_2^b\,{\bf T}_3^b
=if^{bac}{\bf T}_1^a\,{\bf T}_2^c\,{\bf T}_3^b,
\label{colab}
\end{equation}
where ${\bf T}_i^a$ represents a colour generator with adjoint index
$a$ on line $i$, with path ordering outwards from the hard interaction
as indicated in figure~\ref{fig:2loop}. On the right-hand side we have
used the SU(3) colour algebra on line $i$:
\begin{equation}
\left[{\bf T}_i^b,{\bf T}_i^c\right]=if^{bac}{\bf T}_i^c,
\label{SU(3)}
\end{equation}
and we see that the surviving colour factor is directly proportional
to that of figure~\ref{fig:connected}, which is indeed a connected
gluon graph. It should be stressed that it is only the colour factors
that are related in this way: in a general gauge, the kinematic
combination appearing in eq.~(\ref{expabconn}) is completely different
to the kinematic factor of the diagram in
figure~\ref{fig:connected}. Nevertheless, this property is a
generalisation of a similar property observed for the two-line webs of
refs.~\cite{Gatheral:1983cz,Frenkel:1984pz,Sterman:1981jc}, whose
colour factors were found to be ``maximally non-abelian'', namely
equivalent to fully connected graphs.\\

The fact that only connected colour factors appear in the exponent of
the soft function was proven completely generally in
ref.~\cite{Gardi:2013ita}, for all types of web at arbitrary loop
order. To this end, an alternative formalism to the use of web mixing
matrices was set up, which involved rewriting replica-ordered Wilson
products as single exponential factors, using the
Baker-Campbell-Hausdorff formula. Although web mixing matrices do not
appear explicitly in this approach, the price one pays is that higher
order webs must be calculated with an increasing number of effective
vertices describing emission of multiple gluons from each Wilson
line. These calculations become cumbersome quickly as the loop order
increases. However, the effective vertex formalism provides a
well-motivated choice of basis for the connected colour factors
appearing at each order~\cite{Gardi:2013ita}. Recently, an alternative
approach to calculating soft exponents has appeared, which also
involves effective
vertices~\cite{Vladimirov:2014wga,Vladimirov:2015fea}. Interestingly,
in this approach the replica trick is not used at all.\\

The classification of web mixing matrices, and calculation of their
associated colour factors, is only part of the story of understanding
webs. As eq.~(\ref{Sstruc}) makes clear, in order for webs to be
useful, one must also calculate the kinematic factors associated with
the web Feynman diagrams themselves. This is the subject of the
following section.

\section{Calculation of webs}
\label{sec:renorm}

As we have seen, web diagrams are Feynman diagrams involving Wilson
lines. These can be calculated in either position or momentum space,
and the technology needed to perform such calculations has developed
over many years (see
e.g.~\cite{Korchemsky:1985xj,Korchemsky:1987wg,Korchemsky:1985xu,Brandt:1981kf,Aybat:2006wq,Aybat:2006mz,Becher:2009cu,
  Gardi:2009qi,Becher:2009qa,Kidonakis:2009ev,Becher:2009kw,Ferroglia:2009ii,
  Ferroglia:2009ep,Mitov:2009sv,Gardi:2013saa,Falcioni:2014pka,Laenen:2015jia,Laenen:2014jga,Grozin:2014hna,Grozin:2014axa,Henn:2013wfa,Henn:2012qz,Correa:2012nk,Correa:2012at}).
The current state of the art is two-loops in full multiparton
scattering (for both massless and massive Wilson lines), and
three-loops for the case of two Wilson lines. \\

\begin{figure}
\begin{center}
\scalebox{0.6}{\includegraphics{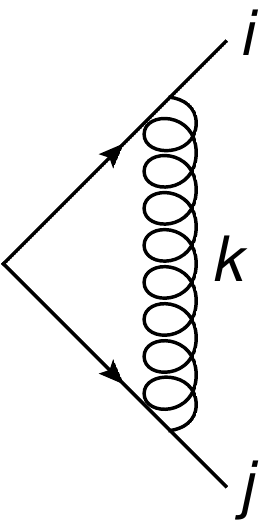}}
\caption{One-loop diagram involving two Wilson lines.}
\label{fig:oneloopcusp}
\end{center}
\end{figure}

Wilson line diagrams contain both IR and UV divergences, corresponding
to taking gluon subgraphs to infinity, or shrinking them to the
origin, respectively. One must then regulate these divergences, and it
is conventional to use dimensional
regularisation~\cite{'tHooft:1972fi} in $4-2\epsilon$
dimensions. Here, however, a complication arises: it turns out that,
for all Wilson line diagrams at arbitrary loop order, the relevant
kinematic integrals vanish. We can see physically why this is by
appealing to a one-loop example, namely the diagram of
figure~\ref{fig:oneloopcusp}, which may represent two out of many
coloured particles coming from a given hard interaction.
Labelling the momentum of line $i$ by $p_i$, the kinematic part of the
diagram is (in momentum space), proportional to
\begin{equation}
\int\frac{d^d k}{(2\pi)^d}\left(\frac{p_i^\mu}{p_i\cdot k}\right)
\left(-\frac{\eta_{\mu\nu}}{k^2}\right)
\left(-\frac{p_j^\mu}{p_j\cdot k}\right)
=\int\frac{d^d k}{(2\pi)^d}\frac{p_i\cdot p_j}{k^2\,p_i\cdot k\,p_j\cdot k},
\label{oneloopgraph}
\end{equation}
where we have used the eikonal Feynman rule of eq.~(\ref{eikrule}),
and the Feynman gauge gluon propagator. As we saw in
section~\ref{sec:abelian}, such Feynman rules occur when simplifying
the momentum space Feynman rules for whole amplitudes (i.e. with
arbitrary $k$). In particular, we have linearised the denominators
associated with the Wilson lines. If they were the usual denominators
associated with legs of momentum $p_i$, eq.~(\ref{oneloopgraph}) would
instead read
\begin{equation}
\int\frac{d^d k}{(2\pi)^d}\frac{p_i\cdot p_j}{k^2\,[(p_i+k)^2-m^2]\,[(p_j-k)^2-m^2]},
\label{oneloopgraph2}
\end{equation}
where we have taken the hard particles to be massive for full
generality.  One may examine the divergence properties of this
integral using na\"{i}ve power counting. This essentially means that
we identify $k$ with some momentum scale $\Lambda$ and see how the
integrand behaves as $\Lambda$ goes to zero (the IR limit) or infinity
(the UV limit), ignoring the fact that all quantities are 4-vectors
etc. One then finds that eq.~(\ref{oneloopgraph}) goes like
$\Lambda^{d-4}$ and $\Lambda^{d-6}$ in the IR and UV regimes
respectively. This tells us that the integral is logarithmically
divergent in $\Lambda$ in 4 spacetime dimensions in the IR regime,
which would show up as a pole in $\epsilon$ in dimensional
regularisation. On the other hand, there is no UV divergence, due to
the terms in the denominators which are quadratic in $k^2$. \\

One finds a different result for the linearised integral of
eq.~(\ref{oneloopgraph}), which is logarithmically divergent (in
$\Lambda$) in both the IR and UV regimes. Comparing the integrals, it
is clear why this is: by linearising the denominators associated with
the Wilson lines, one removes the $k^2$ terms that damp the UV
singularity in eq.~(\ref{oneloopgraph2}). In other words, by taking
the soft limit we have introduced a {\it spurious} UV singularity. A
careful treatment of the integral in eq.~(\ref{oneloopgraph}) shows
that these singularities precisely cancel each other, leaving a zero
result. Furthermore, this phenomenon generalises to all Wilson line
integrals. In other words, the UV singularities of Wilson lines are in
one-to-one correspondence with the IR singularities of
amplitudes~\cite{Korchemsky:1987wg}. It follows that we can extract
physical information from Wilson line integrals by isolating their UV
poles, and ignoring their IR singularities. This can be done using
suitable regulators, over and above the use of dimensional
regularisation. A particularly convenient regulator has been defined
in refs.~\cite{Gardi:2011yz,Gardi:2013saa,Falcioni:2014pka}, and
consists of replacing the usual QCD Wilson line with
\begin{equation}
  \Phi_{\beta_i}^{(m)} \, = \, {\cal P} \exp \left[ {\rm i} g \mu^\epsilon
  \int_0^\infty d \lambda \, \beta_i \cdot A \left( \lambda \beta_i \right) \,
  {\rm e}^{- {\rm i} m \lambda \sqrt{\beta_i^2 - {\rm i} \varepsilon}}
  \right] \, ,
\label{Phidef}
\end{equation}
where we have shown the dependence on the coupling $g$, which the
dimensional regularisation scale $\mu$ ensures is dimensionless. Also,
$\beta_i$ is the 4-velocity of the Wilson line, which is assumed to
have been tilted off the lightcone if null. Equation~(\ref{Phidef})
differs from the usual Wilson line due to an additional exponential
factor, which smoothly cuts off the long-distance contribution to
leave only UV poles in $\epsilon$. One recovers the conventional QCD
result as $m\rightarrow 0$.\\

The structure of UV singularities are encoded in a quantity known as
the {\it soft anomalous dimension}. To define this, let us first
extract the UV poles of the soft function to define a {\it
  renormalised} soft function, that is UV finite:
\begin{equation}
  {\cal S}_{\rm ren.} \left( \gamma_{ij}, \alpha_s(\mu^2), \epsilon, \frac{m}{\mu}
  \right) \, = \, {\cal S} \left( \gamma_{ij}, \alpha_s(\mu^2), \epsilon, \frac{m}{\mu}
  \right) \, Z \left( \gamma_{ij}, \alpha_s(\mu^2), \epsilon \right) \, ,
\label{Srendef}
\end{equation}
We have here shown the dependence of each function on the regulators
$\epsilon$ and $m$, the coupling $\alpha_s$ and dimensional
regularisation scale $\mu$, and the parameters
\begin{equation}
\gamma_{ij}=\frac{2\beta_i\cdot\beta_j}{\sqrt{\beta_i^2}\sqrt{\beta_j^2}}
\label{gammaijdef}
\end{equation}
which are related to the hyperbolic angle, in Minkowski space, between
any two Wilson lines. In eq.~(\ref{Srendef}), $Z$ is a multiplicative
factor that collects all UV counterterms~\footnote{The fact that
  Wilson lines renormalise multiplicatively was first shown in
  refs.~\cite{Brandt:1981kf,Polyakov:1980ca,Arefeva:1980zd,Dotsenko:1979wb}.}. Note
that this equation is more complicated than it looks: both ${\cal S}$
and $Z$ are matrix-valued in the space of colour flows (as discussed
in section~\ref{sec:2vsn}), and thus the order of the factors on the
right-hand side of eq.~(\ref{Srendef}) is important. The counterterm
matrix $Z$ satisfies the renormalisation group equation
\begin{equation}
  \mu \frac{d }{d \mu} Z \left( \gamma_{ij}, \alpha_s(\mu^2), \epsilon \right) \, = \, - \,
  Z \left( \gamma_{ij}, \alpha_s(\mu^2), \epsilon \right)
  \Gamma \left(\gamma_{ij}, \alpha_s(\mu^2) \right) \, .
\label{Zeq}
\end{equation}
The factor $\Gamma$ on the right-hand side is the soft anomalous
dimension, and is finite as $\epsilon\rightarrow 0$. Like the
counterterm matrix $Z$, it is matrix-valued in the space of colour
flows.\\

For practical calculational purposes, there is a direct route from the
unrenormalised soft function (in terms of webs) to the soft anomalous
dimension. One may write the former as
\begin{equation}
  {\cal S} \left(\alpha_s, \epsilon \right) \, = \, \exp \left[ \,
    \sum_{n = 1}^\infty \, \sum_{k = - n}^\infty \alpha_s^n \,
    \epsilon^k \, w^{(n, k)} \right] \, ,
\label{Sunren}
\end{equation}
where $w^{(n,k)}$ is the coefficient of $\epsilon^k$ from the sum of
all web contributions at ${\cal O}(\alpha_s^n)$. In terms of these
coefficients, the soft anomalous dimension is given up to three-loop
order by~\cite{Gardi:2011yz}
\begin{eqnarray}
\label{Gamres}
  \Gamma^{(1)} & = & - 2 w^{(1,-1)} \, ,\nonumber \\
  \Gamma^{(2)} & = & - 4 w^{(2,-1)} - 2 \left[ w^{(1,-1)}, w^{(1,0)} \right] \, ,\nonumber \\
  \Gamma^{(3)} & = & - 6 w^{(3,-1)} + \frac{3}{2} b_0 \left[ w^{(1,-1)}, w^{(1,1)} \right]
  + 3 \left[ w^{(1,0)}, w^{(2,-1)} \right] + 3 \left[ w^{(2,0)}, w^{(1,-1)} \right] \nonumber \\
  & & \hspace{-1cm} 
  + \left[ w^{(1,0)}, \left[w^{(1,-1)}, w^{(1,0)} \right] \right] - \left[ w^{(1,-1)}, 
  \left[w^{(1,-1)}, w^{(1,1)} \right] \right] \, ,
\end{eqnarray}
where $\Gamma^{(i)}$ is the coefficient of $\alpha_s^i$ in the soft
anomalous dimension, and $b_0$ the one-loop coefficient of the QCD
$\beta$ function. We see a very general structure, namely that the
coefficients of $\Gamma$ consist of the single pole of the
unrenormalised webs at each given order, dressed by commutators of
lower-order web coefficients. These commutators are non-zero precisely
because webs are matrix-valued in the space of colour flows. When this
complication is absent (such as in two line scattering, or multiparton
scattering in the limit of a large number of colours), these
commutators disappear. This is intimately related to the fact that
webs in two-parton scatteing (in the language of
refs.~\cite{Gatheral:1983cz,Frenkel:1984pz,Sterman:1981jc}) are single
irreducible diagrams. Such diagrams have a single divergence at
arbitrary loop order, such that the soft anomalous dimension (also
known as the cusp anomalous dimension in this case) inherits all of
its information from the coefficient of the single pole in each
web. For multiparton webs, diagrams can instead be reducible, meaning
that they have higher order poles in $\epsilon$. These are then
cancelled by commutators of lower order webs, in a tightly constrained
way.\\

\begin{center}
\parbox{5in}{{\bf Problem 7}. Consider the two-loop web of
  figure~\ref{fig:2loop}. Identify the single connected colour factor
  that this contributes to. Identify also the lower order webs that
  enter the subtraction term for this web, and show that they give rise
  to the same colour factor.}
\end{center}

The combination of an unrenormalised web with its commutator
contributions has been called a {\it subtracted web} in
refs.~\cite{Gardi:2011yz,Gardi:2013saa,Falcioni:2014pka}, and has a
number of special properties. Firstly, dependence on the regulator $m$
used to separate IR and UV singularities cancels only at the level of
subtracted webs. Secondly, the analytic functions of cusp angles that
appear in subtracted webs are much simpler than for unsubtracted
webs. A crucial property in making the web language useful is that
subtraction terms operate on a web-by-web basis. That is, each closed
set of web diagrams interacts with subtraction terms involving lower
order diagrams from the {\it same} set, such that webs do not mix with
each other under renormalisation. It is for this reason that the
identification of webs as closed sets of diagrams is
meaningful. Different webs would still mix with each other, however,
under gauge transformations. Clever gauge choices can perhaps be used
to simplify the calculation of webs at higher orders e.g. the
conformal gauges of ref.~\cite{Chien:2011wz}.\\

Based on the above discussion, the procedure for calculating the soft
anomalous dimension matrix at a given order is as follows:
\begin{enumerate}
\item Choose a basis of independent connected colour factors.
\item For each web, find (using the web mixing matrix or otherwise)
  the combination of kinematic factors accompanying each connected
  colour factor, if present. Note that the fact that only combinations
  are needed as opposed to individual diagrams potentially leads to
  simplifications, due to cancellations between diagrams at the
  integrand level.
\item Calculate each kinematic combination, including all relevant
  commutator subtraction terms. Further simplifications occur at this
  point, including independence of the regulator used to separate UV
  and IR singularities.
\item Finally, add together all kinematic combinations for a given
  connected colour factor. This gives a gauge-invariant contribution
  to the soft anomalous dimension.
\end{enumerate}
This programme of work is currently being carried out at three-loop
order, where webs may be classified according to the number of three
and four gluon vertices off the Wilson lines. Diagrams with no
multigluon vertices are known as {\it Multiple Gluon Exchange Webs}
(MGEWs). Explicit results have been presented for all three-loop
examples connecting three or more
lines~\cite{Gardi:2013saa,Falcioni:2014pka}, and the two-line case has
been studied in ref.~\cite{Henn:2013wfa}. Certain diagrams of this
type can even be calculated to all orders i.e. for any number of
gluons~\cite{Falcioni:2014pka}. This mirrors parallel developments
(discussed above) in the combinatoric understanding of web mixing
matrices, and suggests that it may be possible to understand all-order
properties in the exponent of the soft amplitude. As an example,
ref.~\cite{Falcioni:2014pka} conjectures a basis of functions which
are believed to be adequate to describe the result of any MGEW, at
arbitrary loop order. This basis, or generalisations thereof, may also
hold for more complicated web types. An important element in the
construction of this basis is the efficient characterisation of
analytic properties using the {\it Symbol} of
refs.~\cite{Goncharov.A.B.:2009tja,Goncharov:2010jf,Duhr:2011zq,Duhr:2012fh},
which tool will doubtless prove to be useful at higher loop
orders. Very recently, the calculation of the fully connected webs
linking four Wilson lines has been completed~\cite{Almelid:2015jia},
in the lightlike limit. As argued by the authors, this is sufficient
to obtain the complete soft anomalous dimension in this limit at
three-loop order. Another interesting recent development is the use of
unitarity-inspired methods to compute
webs~\cite{Laenen:2014jga,Laenen:2015jia}.

\section{Conclusion}
\label{sec:conclude}

In this article, we have provided a review of past and current
developments in the classification of infrared singularities in
scattering amplitudes. These are interesting for a variety of
phenomenological and formal reasons, and we have focused mainly on
the case of non-abelian gauge theories, whose singularity structure is
much richer than QED or gravity. \\

Our main aim has been to provide an accessible introduction to the
language of {\it webs}, namely Feynman diagrams that occur in the
exponent of the soft (infrared divergent) part of an amplitude, itself
a vacuum expectation value of Wilson lines. We have seen that webs in
multiparton scattering are closed sets of diagrams, related by
permutations of gluons on the Wilson lines. The colour and kinematic
parts of these diagrams are entangled by {\it web mixing matrices},
whose mathematical structure makes contact with interesting problems
in contemporary combinatorial research. Further elucidation of these
matrices, from either a physics or mathematics viewpoint, offers
significant insights into the all-order properties of perturbation
theory. This is equally true for the calculation of the kinematic
parts of web diagrams, where systematic methods are beginning to be
developed.\\

Investigation of multiparton webs, and their associated physics in a
variety of gauge theories, is still a very young subject. Many
interesting results are still ripe to be discovered, and work towards
such discoveries is ongoing!

\section*{Acknowledgments}

I am grateful to Stefano Forte for the suggestion and invitation to
publish these notes. My knowledge of this subject has benefited
greatly over the last few years from conversations and collaborations
with Mark Dukes, Giulio Falcioni, Einan Gardi, Mark Harley, Eric
Laenen, Lorenzo Magnea, Heather McAslan, Darren Scott, Jenni Smillie,
Gerben Stavenga and Einar Steingr\'{i}msson. I am supported by the UK
Science and Technology Facilities Council (STFC), under grant
ST/L000446/1.

\bibliography{refs.bib}
\end{document}